\documentclass[12pt]{article}
\usepackage{graphicx}
\usepackage{latexsym,amsmath,amsfonts,amssymb,color}
\usepackage[latin1]{inputenc}
\usepackage{cite}
\usepackage[american]{babel}
\usepackage{hyperref}
\newcommand{\be}{\begin{equation}}
\newcommand{\ee}{\end{equation}}
\newcommand{\beq}{\begin{equation}}
\newcommand{\eeq}{\end{equation}}
\newcommand{\bea}{\begin{eqnarray}}
\newcommand{\eea}{\end{eqnarray}}

\newcommand{\ba}{\begin{eqnarray}}
\newcommand{\ea}{\end{eqnarray}}
\begin{document}
\baselineskip=15.5pt
\pagestyle{plain}
\setcounter{page}{1}

\def\ie{{\em i.e.},}
\def\eg{{\em e.g.},}
\newcommand{\rc}{\nonumber\\}
\newcommand{\bear}{\begin{eqnarray}}
\newcommand{\eear}{\end{eqnarray}}
\newcommand{\Tr}{\mbox{Tr}}    
\newcommand{\ack}[1]{{\color{red}{\bf Pfft!! #1}}}

\numberwithin{equation}{section}

\renewcommand{\theequation}{{\rm\thesection.\arabic{equation}}}


\begin{flushright}
HIP-2016-16/TH
\end{flushright}

\begin{center}

\centerline{\Large {\bf  Quantum phase transitions with dynamical flavors}}

\vspace{8mm}

\renewcommand\thefootnote{\mbox{$\fnsymbol{footnote}$}}
Yago Bea${}^{1,2}$\footnote{yago.bea@fpaxp1.usc.es},
Niko Jokela${}^{3,4}$\footnote{niko.jokela@helsinki.fi},
and Alfonso V. Ramallo${}^{1,2}$\footnote{alfonso@fpaxp1.usc.es}

\vspace{4mm}

\vskip 0.2cm
${}^1${\small \sl Departamento de  F\'\i sica de Part\'\i  culas} \\
{\small \sl Universidade de Santiago de Compostela} \\
{\small \sl and} \\
${}^2${\small \sl Instituto Galego de F\'\i sica de Altas Enerx\'\i as (IGFAE)} \\
{\small \sl E-15782 Santiago de Compostela, Spain} 
\vskip 0.2cm

${}^3${\small \sl Department of Physics} and ${}^4${\small \sl Helsinki Institute of Physics} \\
{\small \sl P.O.Box 64} \\
{\small \sl FIN-00014 University of Helsinki, Finland}

\end{center}

\vspace{8mm}
\numberwithin{equation}{section}
\setcounter{footnote}{0}
\renewcommand\thefootnote{\mbox{\arabic{footnote}}}

\begin{abstract}

We study the properties of a D6-brane probe in the ABJM background with smeared massless dynamical quarks in the Veneziano limit. Working at zero temperature and non-vanishing charge density, we show that the system undergoes a quantum phase transition in which the topology of the brane embedding changes from a black hole to a Minkowski embedding. In the unflavored background the phase transition is of second order and takes place when the charge density vanishes. We determine the corresponding critical exponents and show that the scaling behavior near the quantum critical point has multiplicative logarithmic corrections. In the background with dynamical quarks the phase transition is of first order and occurs at non-zero charge density. In this case we compute the discontinuity of several physical quantities as functions of the number $N_f$ of unquenched quarks of the background. 

\end{abstract}

\newpage
\tableofcontents

\section{Introduction}

Quantum phase transitions are transitions that happen at zero temperature and that are induced by quantum fluctuations. They occur when some control parameters are varied and tuned to critical values, at which the ground state of the system undergoes a macroscopic rearrangement and the energy levels develop a non-analytic behavior on these parameters. Although, the quantum phase transitions occur at zero temperature, they determine the behavior of the system at low temperature in the so-called quantum critical regime, which is a region of the phase diagram surrounding the quantum critical point (see, \eg \cite{Sachdev,Vojta} for reviews).

Strong coupling is a natal environment, where one expects quantum phase transitions. Therefore, a natural question is whether holography could be useful to search and characterize new types of quantum critical matter. Indeed, it is extremely important to develop new theoretical models which could shed light on the nature of quantum criticality and could serve to establish new paradigms to describe these phenomena. 

In the recent years different holographic models displaying quantum phase transitions have been studied in the literature (see, for example \cite{Karch:2007br,Jensen:2010vd,Jensen:2010ga,Evans:2010iy,Iqbal:2010eh,Iqbal:2011aj,D'Hoker:2012ih,Filev:2014mwa}). We are especially interested in top-down models, for which the field theory dual is clearly identified. In particular, we will deal with probe flavor D-branes in a gravitational background,  that corresponds, in the field theory side, to adding fields on the fundamental representation of the gauge group which act as charge carriers. When $N_f$ flavor D-branes are added to a geometry generated by $N_c$ color branes with $N_f\ll N_c$, we can use the probe approximation and neglect the backreaction of the flavor branes on the geometry. 
This precludes the fundamentals being dynamical and they are treated as quenched in the field theory.

The worldvolume dynamics of the flavor branes is governed by an action which has two pieces. The first one is the standard Dirac-Born-Infeld (DBI) action, which contains a gauge field. The other one is the Wess-Zumino (WZ) action which couples the brane to the Ramond-Ramond potentials of the background. The effects from the latter typically lead to far reaching consequences. In this probe brane setup it is rather simple to generate a configuration dual to a compressible state with non-zero charge density \cite{Apreda:2005yz,Kim:2006gp,Horigome:2006xu,Parnachev:2006ev,Kobayashi:2006sb}. Indeed, the charge density is dual to a radial electric field on the worldvolume. When the density is non-vanishing all consistent embeddings reach the horizon, \ie\  are black hole embeddings, whereas at zero  density there could also be Minkowski embeddings which always stay outside the horizon.\footnote{Suitable WZ terms would allow regular Minkowski embeddings even at non-zero charge density \cite{Bergman:2010gm,Jokela:2011eb}.}  It was shown in \cite{Karch:2007br} for the D3-D7 and D3-D5 systems that a quantum phase transition takes place at zero temperature at the point where the charge density vanishes, which corresponds to the chemical potential being equal to the quark mass. This phase transition is of second order and is realized in the holographic dual as a topology change of the embedding (from the black hole to Minkowski).  In \cite{Ammon:2012je} the critical exponents of the transition were found, corresponding to a non-relativistic scale invariant field theory with hyperscaling violation. These results were generalized in \cite{Itsios:2016ffv} to generic D$p$-D$(p+4)$ and D$p$-D$(p+2)$ intersections. 

Our aim is to study the quantum phase transitions of brane probes in the gravity dual of the ABJM Chern-Simons matter theory, especially in the Veneziano limit. This is an $U(N)\times U(N)$ Chern-Simons gauge theory in $2+1$ dimensions with levels $(k,-k)$ and bifundamental fields transforming in the $(N, \bar N )$ and $(\bar N,  N )$ representations of the gauge group. This theory was proposed in \cite{Aharony:2008ug} as the low energy theory of $N$ coincident M2-branes at a ${\mathbb C}^4/{\mathbb Z}_k$ singularity.  When $N$ and $k$ are large the theory admits a supergravity description in the ten-dimensional type IIA theory. The corresponding geometry is of the form $AdS_4\times {\mathbb C}{\mathbb P}^3$ with fluxes (see \cite{Klebanov:2009sg,Klose:2010ki,Marino:2011nm,Bagger:2012jb} for reviews of several aspects of the ABJM model).

The flavors in the ABJM theory are fields transforming in the fundamental representations $(N,1)$ and $(1,N)$ of the gauge group. In the holographic dual these flavors are introduced by means of D6-branes extended in $AdS_4$ and wrapping an ${\mathbb R}{\mathbb P}^3$ cycle inside the ${\mathbb C}{\mathbb P}^3$ internal manifold
\cite{Gaiotto:2009tk,Hohenegger:2009as}. In the probe approximation these holographic quarks have been studied in  \cite{Hikida:2009tp,Jensen:2010vx,Ammon:2009wc, Alanen:2009cn, Zafrir:2012yg}. Moreover, by using the smearing technique when $N_f$ is large,  one can obtain simple analytic geometries encoding the effects of dynamical quarks in holography (see \cite{Nunez:2010sf} for a review of this general method).

In general, in order to obtain the gravity dual of a field theory with unquenched flavor, one has to solve the equations of motion of supergravity with brane sources. If the flavor branes are localized the sources have Dirac $\delta$-funcions and the problem of solving the equations of motion is extremely difficult. In the smearing approach this difficulty is overcome by considering a continuous distribution of flavor branes in the internal space. In many cases this simplification allows to find simple solutions of the equations of motion of the gravity-plus-branes system.  The price one has to pay for this simplification is the modification of the field theory dual. First of all, the amount of supersymmetry preserved by the smearing background is less than the one preserved by the localized setup. Moreover, the smeared flavor branes are not coincident and, therefore, the flavor symmetry for $N_f$ flavors is $U(1)^{N_f}$ rather than $U(N_f)$. Finally, the fact that we are superimposing branes with different orientations implies that we are modifying the R-symmetry of the theory.

The geometry generated by the backreaction of massless flavors in ABJM has been obtained in \cite{Conde:2011sw} at zero temperature and generalized in \cite{Jokela:2012dw} to non-vanishing temperature. The backreaction affects the ABJM geometry  rather mildly since the metric differs from the unflavored one by constant squashing factors which depend on $N_f$. For massive quarks this construction was carried out in \cite{Bea:2013jxa} (see also \cite{Bea:2014yda}), leading to bigger modifications of the background geometry. 
In the case of the ABJM model with massless flavors, the backreacted geometry is of the form 
$AdS_4\times {\cal M}_6$, where ${\cal M}_6$ is a squashed deformation of ${\mathbb C}{\mathbb P}^3$. Since the massless flavored metric continues to have an Anti-de-Sitter  factor, it is straightforward to find its finite temperature deformation by simply including a blackening factor in the Anti-de-Sitter part, without modifying the internal metric. This is a particular simplification of the ABJM model which does not occur it other theories (see \cite{Bigazzi:2009bk} for the analysis of the D3-D7 black hole background).  In \cite{Jokela:2012dw} it was checked that the non-zero temperature flavored ABJM background gives rise to a consistent thermodynamics  and passes some highly non-trivial consistency tests.

In this paper we probe the ABJM background (with and without massless dynamical quarks included) with a flavor D6-brane corresponding to a massive quark. We study the dynamics of this probe at zero temperature and non-vanishing  charge density. This dynamics is governed by the DBI action, with the WZ term playing a fundamental role. 
We are interested in the phase structure of the system as the charge density is varied and, in particular, in analyzing the phase transition that occurs when the charge density is small.

We first study the probe in the unflavored ABJM background. Working at zero temperature, we find a continuous quantum phase transition at the point where the charge density vanishes. This transition is similar to the one that happens in the D$p$-D$q$ systems  in \cite{Ammon:2012je,Itsios:2016ffv} and corresponds to passing from a black hole to a Minkowski embedding. However, the scaling behavior of the probe near the critical point differs from the ones found in \cite{Ammon:2012je,Itsios:2016ffv}. Indeed, we find that the corresponding critical exponents are different and, in addition, our system displays multiplicative logarithmic corrections to the scaling behavior.

\begin{figure}[ht]
\center
 \includegraphics[width=0.50\textwidth]{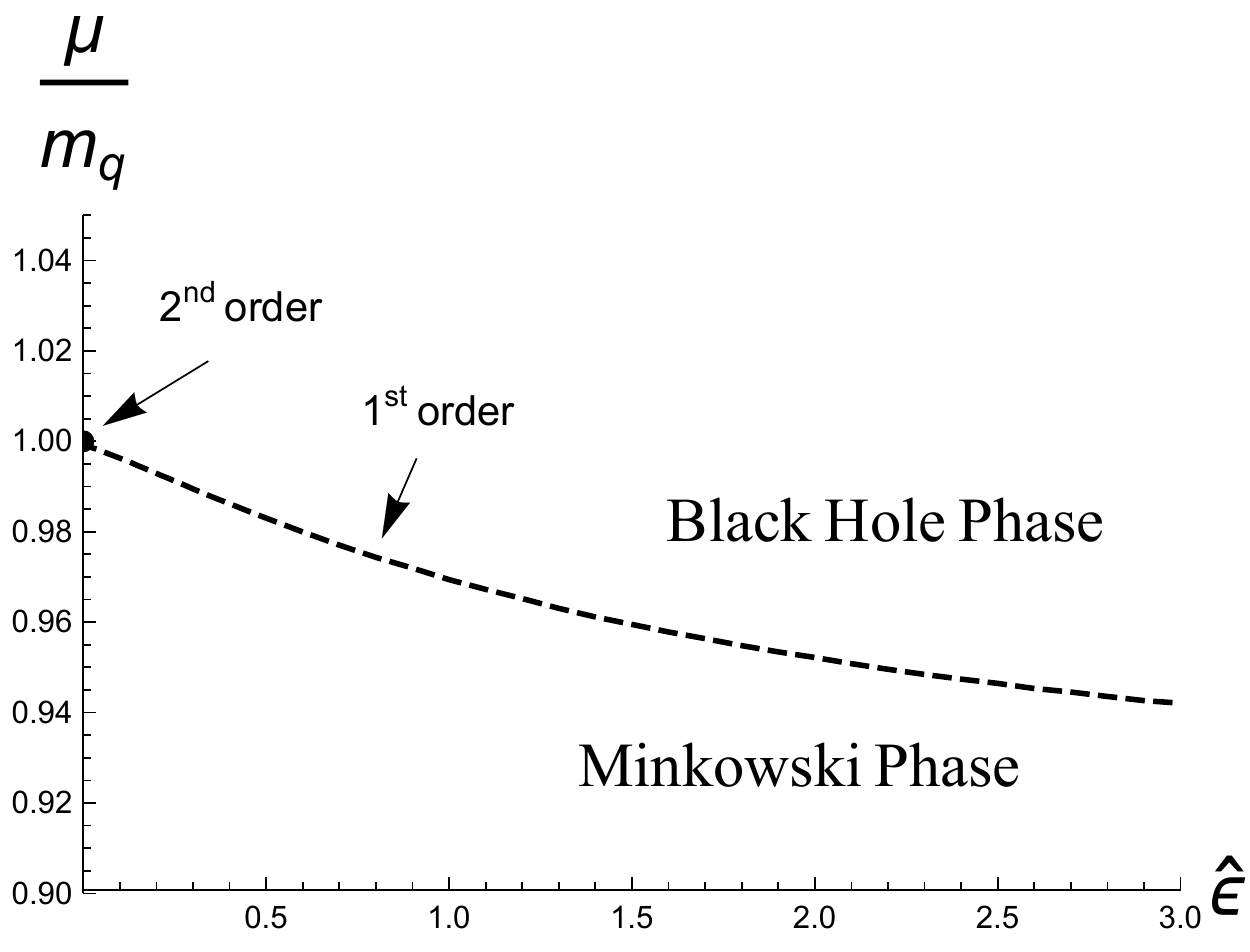}
  \caption{The phase diagram of the unquenched  ABJM model at zero temperature separates two different domains. At high enough chemical potential to quark mass ratios, the system is in the so-called black hole phase, which corresponds to a metallic behavior. The lower domain stands for Minkowski phase, where the system is gapped to charged excitations and resembles an insulating phase. The two domains are separated by a curve of first order phase transitions, whose location depends on the amount of flavor in the background $\hat \epsilon\propto N_f$ (see (\ref{hatepsilon})) for a given chemical potential. The curve ends at the second order critical point in the quenched limit $N_f\to 0$. Interestingly, the corresponding critical exponents characterizing the continuous phase transition exhibit multiplicative logarithmic corrections.} 
  \label{fig:phasediagram}
\end{figure}

We also study the effects due to the presence of unquenched dynamical quarks in the background. In general, the inclusion of the flavor backreaction in holography is quite challenging. However, in the ABJM model the deformation of the geometry due to massless flavors seems quite mild and this gives us a unique opportunity to explore the different flavor effects. What we found below is that the influence on the phase transition of the unquenched case is not so moderate as their effects of the geometry could suggest. Indeed, we show below that the flavored black hole to Minkowski phase transition occurs at non-zero density and, moreover, it is of first order. The phase diagram  at zero temperature is summarized in Fig.~\ref{fig:phasediagram}. We have been able to compute several quantities characterizing this discontinuous transition, such as its latent heat and the speed of sound close to the transition point.

The rest of this paper is organized as follows. In section \ref{background} we review the ABJM background with unquenched flavor. In section \ref{Probes} we study the embeddings of flavor D6-branes, both at zero and non-zero temperature. Section \ref{zeroT_thermo} is devoted to the analysis of the zero temperature thermodynamics and to explore  the quantum phase transitions in the unflavored and flavored cases. In section \ref{diffusion} we determine the charge susceptibility and diffusion constants at non-zero temperature. In section \ref{fluctuations} we analyze the fluctuations of the probe  and, in particular, we calculate the speed of its zero sound mode. In section \ref{summary} we summarize our results and discuss possible future research directions.  The paper is completed with two appendices. In appendix \ref{background_appendix} we give further details of the flavored background and of the embeddings of the probes. Finally, in appendix \ref{fluctuation_apendix} we carry out in detail the analysis of the fluctuations of the D6-brane.

\section{The flavored ABJM background}
\label{background}

Let us review the geometry of the ABJM model with smeared massless flavors at non-zero temperature \cite{Conde:2011sw,Jokela:2012dw}. Further details are provided in appendix \ref{background_appendix}. The ten-dimensional metric, in string frame, has the form:
\beq
ds^2\,=\,L^2\,\,ds^2_{BH_4}\,+\,ds^2_{6}\,\,,
\label{flavoredBH-metric}
\eeq
where $L$ is constant (the radius of curvature), $ds^2_{BH_4}$ is the metric of a black hole in four-dimensional Anti-de Sitter, given by:
\beq
 ds^2_{BH_4} = -r^2h(r) dt^2+\frac{dr^2}{r^2h(r)}+r^2\big[dx^2+dy^2\big] \ ,
 \label{BH4-metric}
\eeq
and $ds^2_{6}$ is the metric of the compact six-dimensional manifold. In (\ref{BH4-metric}) the function $h(r)$ is the blackening factor:
\beq
h(r)\,=\,1\,-\,\frac{r_h^3}{r^3} \ ,
\label{blackening-factor}
\eeq
where $r_h$ is the horizon radius, which is proportional to the temperature as $T=3r_h/4\pi$. The six-dimensional internal metric $ds^2_{6}$ in (\ref{flavoredBH-metric})  can be written as an ${\mathbb S}^2$-bundle over  ${\mathbb S}^4$. If 
$ds^2_{{\mathbb S}^4}$ is the standard metric of the unit round ${\mathbb S}^4$ and $z^i$ ($i=1,2,3$),  with 
$\sum_i (z^i)^2\,=\,1$, are the components of a unit three-vector which parameterize a unit two-sphere, then the line element $ds^2_{6}$ is:
\beq
ds^2_{6}\,=\,{L^2\over b^2}\,\,\Big[\,
q\,ds^2_{{\mathbb S}^4}\,+\,\big(d z^i\,+\, \epsilon^{ijk}\,A^j\,z^k\,\big)^2\,\Big] \ .
\label{internal-metric-flavored}
\eeq
Here $A^i$ are the components of the non-abelian one-form connection corresponding to an $SU(2)$ instanton in 
${\mathbb S}^4$, $b$ and $q$ are constant squashing factors which depend on the numbers of flavors and colors  ($N_f$ and $N$) and on the Chern-Simons level $k$ through the combination:
\beq
\hat \epsilon\,\equiv\,{3N_f\over 4k}\,=\,{3\over 4}\,\,{N_f\over N}\,\lambda \ .
\label{hatepsilon}
\eeq
The factor $3/4$ is introduced for convenience. In the last step we have introduced the 't Hooft coupling $\lambda=N/k$. In terms of the deformation parameter $\hat\epsilon$, the squashing factors $q$ and $b$ are:
\bea
 q & = & 3+\frac{3}{2}\hat\epsilon -2\sqrt{1+\hat\epsilon+\frac{9}{16}\hat\epsilon^2} \rc
 b & = & \frac{2q}{q+1}\ .\label{q-b}
\eea
Notice that in the unflavored ABJM background $\hat \epsilon=0$ and $b=q=1$. In this case the internal metric (\ref{internal-metric-flavored})  becomes the canonical Fubini-Study metric of ${\mathbb{CP}}^3$ with radius $2L$ in the so-called twistor representation, and (\ref{flavoredBH-metric}) is the line element  of the ABJM model at non-zero temperature without flavors.  The backreaction of the delocalized D6-brane sources deforms the internal metric by squashing the ${\mathbb{CP}}^3$ relative to the $AdS_4$, and deforms internally the 
${\mathbb{CP}}^3$,  preserving the ${\mathbb S}^4-{\mathbb S}^2$ split. These flavor deformations are encoded in the $q$ and $b$ parameters.  As functions of $\hat \epsilon$,  $q$ and $b$ are monotonically  increasing functions, which approach the values $q\to 5/3$ and $b\to 5/4$ as 
$\hat\epsilon\to\infty$. Moreover, the $AdS_4$ radius $L$ also depends on $\hat \epsilon$ as:
\beq
L^2\,=\,\pi\sqrt{2\lambda}\,
\sqrt{{(4-3b)(2-b)b^3\over 2(b-1)(1+\hat\epsilon)+b}} \,\,.
\label{AdS_radius}
\eeq
Notice that $L^2=\pi\,\sqrt{2\lambda}$ in the unflavored geometry, which means that the flavor effects on $L$  are contained in the second square root in (\ref{AdS_radius}).  The complete solution of type IIA supergravity with sources is endowed with RR two- and four-forms $F_2$ and $F_4$, as well as with a constant dilaton $\phi$ (whose value depends on $N$, $N_f$, and $k$). Their explicit expressions are given in appendix \ref{background_appendix}.

\section{Probes on the flavored ABJM }
\label{Probes}

We are interested in analyzing the behavior of a flavor D6-brane probe in the background described in section \ref{background}.  This flavor brane is extended along the $AdS_4$ coordinates $(x^{\mu}, r)$ and wraps a compact three-dimensional submanifold of the internal space. The precise embedding of this submanifold in the flavored squashed ${\mathbb C}{\mathbb P}^3$ can be found in appendix \ref{background_appendix}.  The induced metric on the worldvolume of the  flavor D6-brane is:
\bear
 {ds^2_{7}\over L^2} &=& r^2\,\left[-h(r)\,dt^2+
 dx^2+dy^2\right]+
  {1\over r^2 }\,
\left({1\over h(r)}+{r^2\,\theta'^{\,2}\over b^2}\,\right)\,dr^2 \rc\rc
&&\qquad +\,{1\over b^2}\,\Big[
q\,d\alpha^2+q\,\sin^2\alpha \,d\beta^2+ \sin^2\,\theta\,\left(\,d\psi\,+\,\cos\alpha\,d\beta\,\right)^2\,\Big] \ ,
\label{induced_metric}
\eear
where $\alpha$, $\beta$, and $\psi$ are angles taking values in the range 
$0\le \alpha < \pi$, $0\le \beta, \psi<2\pi$, and $\theta=\theta(r)$ is an angle which determines the profile of the probe brane.  We want to deal with a system with non-zero baryonic charge density. Therefore, we should have a non-zero value of the $tr$ component of the worldvolume gauge field strength $F=dA$. Accordingly, we will adopt the  following  ansatz:
\beq
\theta\,=\,\theta(r)\,\,,
\qquad\qquad
A\,=\,L^2\,A_t(r)\,dt\,\,.
\label{unperturbed_theta_A}
\eeq
The D6-brane probe is governed by the standard DBI+WZ action:
\beq
S\,=\,S_{DBI}\,+\,S_{WZ}\,\,,
\eeq
where $S_{DBI}$ and $S_{WZ}$ are given by:
\bear
S_{DBI} & = & -T_{D6}\,\int_{{\cal M}_7}\,d^7\zeta\,e^{-\phi}\,
\sqrt{-\det (g+F)}\rc
S_{WZ} & = & T_{D6}\int_{{\cal M}_7}\,\left(\hat C_7\,+\,\hat C_5\wedge F\,+\,{1\over 2}\,
\hat C_3\wedge F\wedge F\,+\,{1\over 6}\hat C_1\wedge F\wedge F\wedge F \right) \ .
\label{DBI_Wess_Zumino}
\eear
In (\ref{DBI_Wess_Zumino})   $g$ is the  induced metric on the worldvolume and the $\hat C_p$'s are the pullbacks of the different RR potentials of the background. In the flavored ABJM background $dF_2\not=0$ and, therefore, the RR potential $C_1$ is not well-defined. In this unquenched case one should work directly with the equations of motion of the probe derived from $S$, which contain the RR field strengths $F_p$ (and do not contain the potentials) (see \cite{Bea:2014yda}). Nevertheless, to determine the embedding corresponding to the ansatz (\ref{unperturbed_theta_A}), only the term with $C_7$ in (\ref{DBI_Wess_Zumino}) is relevant (the explicit expression of $C_7$ can be found in \cite{Conde:2011sw,Jokela:2012dw}).

We will use the following system of worldvolume coordinates 
$\zeta^{a}\,=\,(x^{\mu}, r, \alpha, \beta, \psi)$. After integrating over the internal coordinates, we can write the action in the form:
\beq
S\,=\,\int d^3x\,dr\,{\cal L}\,\,,
\eeq
where ${\cal L}$ is the Lagrangian density of the probe, given by:
\beq
{\cal L}\,=\,-{\cal N}\,r^2\,\sin\theta\,
\Big[\sqrt{b^2(1-A_t'^{\,2})+r^2\,h\,\theta'^{\,2}}-b\sin\theta\,-\,r\,\cos\theta\,\theta'\Big] \ .
\label{total_lag_density}
\eeq
Here and in the following, the prime denotes differentiation with respect to $r$. In (\ref{total_lag_density})   ${\cal N}$ is  a constant given by:
\beq
{\cal N}\,=\,{8\pi^2\,L^7\,T_{D6}\,e^{-\phi}\over b^4}\,q\,\,.
\label{calN_def}
\eeq
In the Lagrangian (\ref{total_lag_density})
the variable $A_t$ is cyclic and its equation of motion can be integrated once as:
\beq
r^2\,\sin\theta\,{A_t'\over 
\sqrt{1-A_t'^{\,2}+{r^2\over b^2}\,h\,\theta'^{\,2}}}\,=\,d\,\,,
\eeq
where $d$ is a constant, which is proportional to the charge density. This equation can be inverted to give:
\beq
A_t'\,=\, {d\over b}{\sqrt{b^2+r^2\,h\,\theta'^{\,2}}\over \,\sqrt{d^2+r^4\sin^2\theta}}\,\,.
\label{At_prime}
\eeq
According to the standard AdS/CFT dictionary  the chemical potential $\mu$ is identified with the value of $A_t$ at the UV:
\beq
\mu\,=\,A_t(r\to\infty)\,\,.
\label{mu_At_UV}
\eeq
For a black hole embedding one can write an expression for $A_t$ as an integral over the radial variable $r$. Indeed, in this case we integrate (\ref{At_prime}) with the condition $A_t(r=r_h)=0$, namely:
\beq
A_t(r)\,=\,{d\over b}\,\int_{r_h}^{r}\,{\sqrt{b^2+\tilde r^2\,h\,\theta'^{\,2}}\over \,\sqrt{d^2+\tilde r^4\sin^2\theta}}
\,\,d\tilde r\,\,.
\eeq
Then, it follows that the chemical potential $\mu$ for a black hole embedding is:
\beq
\mu\,=\,{d\over b}\,\int_{r_h}^{\infty}\,{\sqrt{b^2+r^2\,h\,\theta'^{\,2}}\over \,\sqrt{d^2+r^4\sin^2\theta}}
\,dr\,\,.
\label{chemical_pot}
\eeq

Let us now write the equation of motion for $\theta(r)$:
\beq
\partial_r\,\Bigg[
{r^4\,h\,\sin\theta\over \sqrt{1-A_t'^{\,2}+{r^2\over b^2}\,h\,\theta'^{\,2}}}\,\,\theta'\Bigg]\,-\,
b\,r^2\,\cos\theta\Bigg[(3-2b)\sin\theta\,+\,b\,\sqrt{1-A_t'^{\,2}+{r^2\over b^2}\,h\,\theta'^{\,2}}
\,\Bigg]\,=\,0\,\,.
\label{eom_theta_At}
\eeq
Using (\ref{At_prime}) to eliminate $A_t'$, we can rewrite (\ref{eom_theta_At}) as:
\beq
\partial_r\,\Bigg[ 
{r^2\,h\,\sqrt{d^2+r^4\sin^2\theta}\over 
\sqrt{b^2+r^2\,h\,\theta'^{\,2}}}\,\theta'\Bigg]\,-\,\,r^2\cos\theta\sin\theta\,
\Bigg[3-2b+{r^2\,\sqrt{b^2+r^2\,h\,\theta'^{\,2}}\over
\sqrt{d^2+r^4\sin^2\theta}}\Bigg]\,=\,0\,\,.
\label{eom_theta_d}
\eeq

Eq. (\ref{eom_theta_d}) must be solved numerically, except in the case of vanishing temperature and density, where an analytic supersymmetric solution is available \cite{Conde:2011sw}.   All solutions of (\ref{eom_theta_d}) reach the UV with an angle which approaches asymptotically the value $\theta=\pi/2$. Actually, for large $r$ the deviation of $\theta$ with respect to this asymptotic value can be represented as:
\beq
{\pi\over 2}-\theta(r)\sim {m\over r^b}\,+\,{c\over r^{3-b}}\,+\,\cdots\,\,,
\label{asymptotic}
\eeq
where $b$ is the constant (depending of the flavor deformation parameter $\hat\epsilon$) defined in (\ref{q-b}) and $m$ and $c$ are constants related to the quark mass and the condensate, respectively. The precise holographic dictionary for our probes has been worked out in \cite{Jokela:2012dw}. For our purposes it is sufficient to recall that the physical quark mass $m_q$ is proportional to $m^{{1\over b}}$. This non-trivial exponent is related to the anomalous mass dimension $\gamma_m=b-1$, which enters in the (holographic) Callan-Zymanzik equation \cite{Jokela:2013qya}.  

\begin{figure}[ht]
\center
 \includegraphics[width=0.60\textwidth]{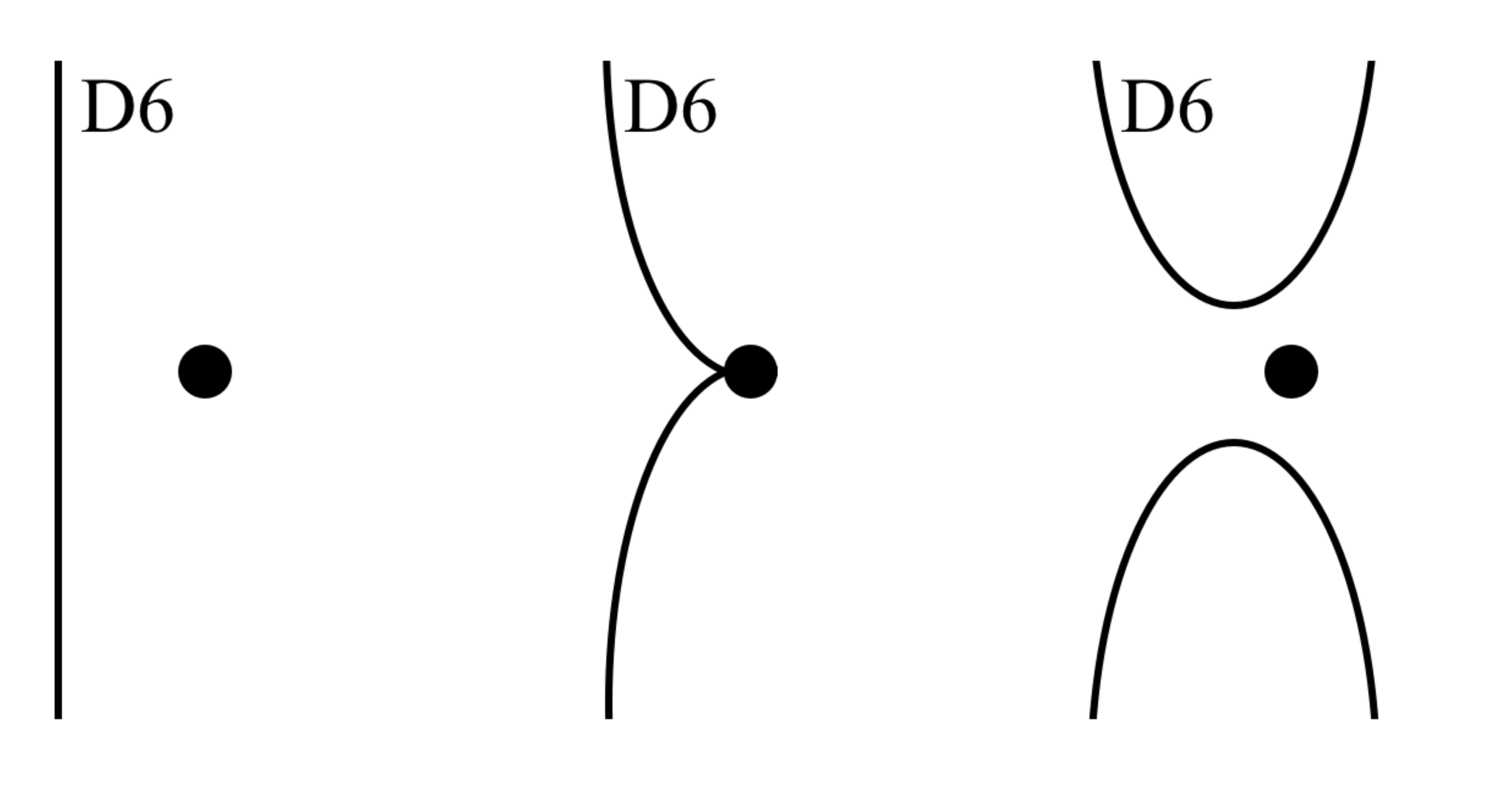}
  \caption{We sketch the three possible embeddings available in the model at non-zero chemical potentials at non-vanishing mass parameter at zero temperature. The left-most profile corresponds to Minkowski embeddings, where the D6-brane does not enter the Poincar\'e horizon, displayed as the black dot. The middle profile corresponds to that of a black hole embedding penetrating the horizon, while the right-most profile stands for D6-anti-D6-brane embeddings. This figure is adapted from the one in \cite{Karch:2007br}, in the context of a D3-D7 model, where a clean flat space interpretation can be given.} 
  \label{embeddings2}
\end{figure}

The different solutions of (\ref{eom_theta_d}) are obtained by imposing suitable boundary conditions  at the  IR. We will study them in the next two subsections, starting with the embeddings at zero temperature.  
There are three different kinds of embeddings, sketched in Fig.~\ref{embeddings2}. They are introduced one-by-one in the following subsection.

\subsection{Embeddings at zero temperature}\label{zeroT_embeddings}

Let us now consider eq. (\ref{eom_theta_d}) for $T=0$ (\ie\ for $h=1$). One can verify by numerical integration that  (\ref{eom_theta_d}) admits a family of solutions in which the embeddings reach the origin $r=0$ at any given value of $\theta_0=\theta(r=0)$, quantities which we shall denote as initial angles.  These solutions are called black hole embeddings as they are continuously connected with their $T\ne 0$ counterparts.  Actually, one can solve (\ref{eom_theta_d}) for $h=1$ in a power series expansion near $r=0$ as:
\beq
\theta(r)=\theta_0\,+\,b(3-2b)\,{\sin\theta_0\cos\theta_0\over 6d}\,r^2\,+\,\cdots\,\,.
\eeq
These solutions can be found numerically by imposing the initial conditions $\theta(r=0)=\theta_0$ and $\theta'(r=0)=0$. The mass parameter $m$ of the embedding (determined by the value of $r^b\cos\theta$ at $r\to\infty$) is related to the initial angle $\theta_0$.  Given the embedding, the chemical potential  can be obtained  by evaluating  the integral (\ref{chemical_pot}). When $\theta_0\to \pi/2$ the mass approaches zero. In fact, the whole embedding becomes trivial with constant angle. When $\theta_0\to 0$, on the other hand, the embedding becomes increasingly spiky and the corresponding chemical potential approaches the value:
\beq
\lim_{\theta_0\to 0}\,\mu\,=\,m^{{1\over b}} \ ,
\label{mu_for_zero_theta}
\eeq
where the mass parameter is kept fixed.

\begin{figure}[ht]
\center
 \includegraphics[width=0.50\textwidth]{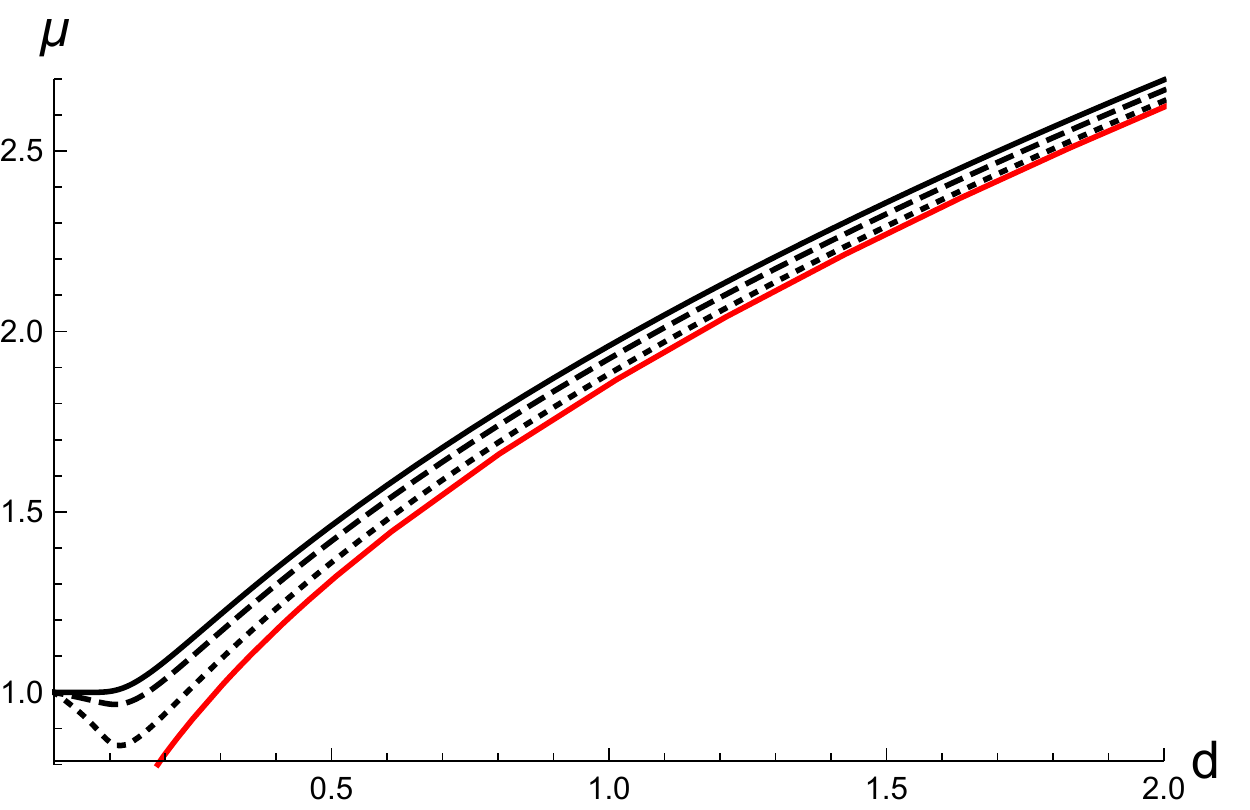}
  \caption{We plot the chemical potential $\mu$ as a function of the density $d$ for fixed quark mass. The continuous black curve corresponds to the unflavored case ($b=1)$, the dashed curve is for $b=1.1$, while the dotted curve is for $b=1.25$ (corresponding to $\hat\epsilon\to\infty$).  All curves are for $m=1$. The  continuous red curve corresponds to the conformal D3-D5 system with massless quarks, for which $\mu=\gamma\, d^{{1\over 2}}$, with 
 $\gamma={1\over 4\sqrt{\pi}}\,\Gamma(1/4)^2$. }
\label{plot_chemical_pot}
\end{figure}

\begin{figure}[ht]
\center
 \includegraphics[width=0.50\textwidth]{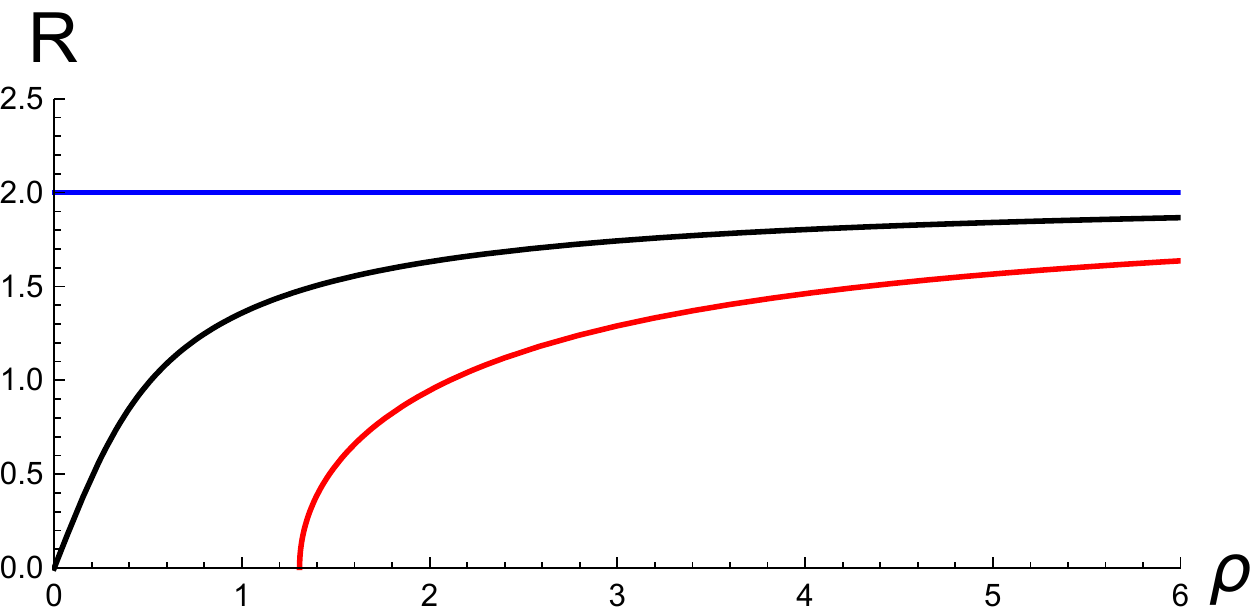}
  \caption{We depict the profiles of the three possible types of embeddings in the $(\rho, R)$ coordinates defined in
  (\ref{R_rho_def}) at zero temperature. The top-most curve corresponds to the Minkowski embedding, while the middle curve entering in the Poincar\'e horizon corresponds to the black hole embedding. The bottom curve stands for one of the branches of the brane-antibrane embeddings.} 
  \label{embeddings}
\end{figure}

We have verified the limit in (\ref{mu_for_zero_theta}) numerically. This result can also be easily demonstrated analytically as follows. Let us first introduce the Cartesian-like coordinates $(\rho, R)$, related to $(\theta, r)$ as:
\beq
R\,=\,r^{b}\,\cos\theta\,\,,
\qquad\qquad
\rho\,=\,r^{b}\,\sin\theta\,\,.
\label{R_rho_def}
\eeq
In these coordinates the black hole embeddings start in the IR at the origin $R=\rho=0$ with a certain angle $\theta_0$ with respect to the $R$-axis and they end at the UV at $R=m$ with $\rho\to\infty$ (see Fig.~\ref{embeddings}).  If the initial angle $\theta_0$ is very small, the embeddings are very spiky and approach the maximal value $R=m$ very fast  for very small values of the coordinate $\rho$. Instead of parameterizing the embedding as $\theta=\theta(r)$, it is more convenient in this situation to represent it as $\rho=\rho(R)$. It is then straightforward to demonstrate that $\mu$ is given by the integral:
\beq
\mu\,=\,{d\over b}\,\int_0^{m}\,
{\sqrt{1+({d\rho\over dR})^2}\over 
\sqrt{d^2(\rho^2+R^2)^{1-{1\over b}}+\rho^2 (\rho^2+R^2)^{{1\over b}}} }\,\,
dR\,\,.
\label{mu_R_rho}
\eeq
For $\theta_0\to 0$ the coordinate $\rho$ is very close to zero except when $R\approx m$ and we can approximate the integral (\ref{mu_R_rho}) by taking $\rho\approx 0$ in the integrand. We get:
\beq
\mu\,\approx\,{1\over b}\,\int_{0}^{m} R^{{1\over b}-1}\,\,dR\,=\,m^{{1\over b}}\,\,,
\eeq
in agreement with (\ref{mu_for_zero_theta}). For a fixed value of the mass parameter $m$, the limiting value (\ref{mu_for_zero_theta})  corresponds to sending $d\to 0$. Actually, the dependence of $\mu$ on $d$ for fixed $m$ can be obtained numerically by performing the integral (\ref{chemical_pot}). The result is shown in Fig.~\ref{plot_chemical_pot}, where we notice an important difference between the unflavored and flavored cases. Indeed, when $N_f=0$ the chemical potential $\mu$ grows monotonically with $d$, starting from its minimal value $\mu=m$ at $d=0$.  When $d$ is large the chemical potential grows as $\mu\propto d^{{1\over 2}}$, which is the behavior expected in a conformal theory in 2+1 dimensions.  On the contrary, when the backreaction  of the flavors is added, $\mu$ decreases for small values of $d$ until it reaches a minimum at a non-zero value of $d$ and then it grows and converges eventually to the unflavored case. The presence of minimum in the $\mu=\mu(d)$ curve means that the charge susceptibility $\chi=\partial d/\partial \mu$ diverges at $d\not=0$, signaling a discontinuous phase transition at a non-zero density. We will confirm this fact below.

The black hole embeddings considered above are not the only possible ones. Indeed, there are also two other configurations in which the brane does not reach the $r=0$ origin. The so-called brane-antibrane embeddings are characterized by the initial boundary conditions:
\beq
\theta(r_0)\,=\,{\pi\over 2}\,\,,
\qquad\qquad
\theta'(r_0)\,=\,\infty\,\,,
\eeq
where $r_0$ is the minimal value of $r$. In terms of the $(\rho, R)$ variables the brane is orthogonal to the $\rho$-axis in the IR (at $\rho=\rho_0=r_0^b$, $R=0$) and becomes parallel to the $\rho$-axis as $\rho$ becomes large (see Fig.~\ref{embeddings}). Notice that $dR/d\rho$ diverges at $\rho=\rho_0$, which indicates that the brane has a turn-around point where the brane jumps to a second branch.

A third class of configurations are the so-called Minkowski embeddings, in which the brane reaches the $R$-axis at some non-zero value of $R$, as shown in Fig.~\ref{embeddings}. Due to charge conservation these embeddings are  only consistent if the density $d$ is zero. When this is the case there are analytic solutions which preserve some amount of supersymmetry \cite{Conde:2011sw}.  In terms of the $(r, \theta)$ variables, these embedding are:
\beq
\cos\theta(r)\,=\,{m\over r^b}\,\,,
\qquad\qquad
(d=0)\,\,.
\label{SUSY_emb}
\eeq
Equivalently $R=m$. Notice that in this case the minimal value of $r$ is $r_0=m^{{1\over b}}$. Moreover, when $d$ vanishes it follows from (\ref{At_prime}) that $A_t'=0$ and, therefore, the gauge field $A_t$ is an arbitrary constant, which equals the chemical potential $\mu$.  Thus, the SUSY embeddings (\ref{SUSY_emb})  correspond to $d=0$, with $\mu$ being a free parameter. 

Notice that, in this zero temperature case, the mass parameter $m$ can be scaled out by a suitable change of the radial variable followed by some redefinitions. Indeed, from (\ref{asymptotic}) we conclude that $m$ can be taken to be one if one changes variables from $r$ to $\tilde r=r/m^{{1\over b}}$. Then, it follows from (\ref{chemical_pot}) that $m$ can be eliminated from this  last equation if $d$ and $\mu$ are written in terms of the rescaled quantities $\tilde d$ and $\tilde \mu$, defined as $\tilde d =d/m^{{2\over b}}$ and  $\tilde\mu=\mu/m^{{1\over b}}$.

In section \ref{zeroT_thermo} we will determine which of these three types of embeddings at zero temperature is thermodynamically favored. We will carry out this analysis by comparing their thermodynamic potentials $\Omega$ in the grand canonical ensemble.

\subsection{Embeddings at finite temperature}\label{finiteT_embeddings}

As will become clear later, we need to extend some of our analysis to small and non-zero temperature. All three types of embeddings, as discussed in the preceding section, extend continuously to $T\ne 0$. However, as our main motivation in this work are the quantum critical phenomena, we will restrict our attention in the black hole phase.
Let us thus only consider the black hole embeddings at non-zero temperature. These embeddings reach the horizon $r=r_h$ with some angle $\theta=\theta_0$. Near $r=r_h$ we can solve (\ref{eom_theta_d}) in powers of $r-r_h$. The first two terms in this expansion are:
\beq
\theta(r)\,=\,\theta_0\,+\,\theta_1\,(r-r_h)\,+\,\cdots\,\,,
\label{expansion_theta_nh}
\eeq
where the constant $\theta_1$ is given by:
\beq
\theta_1\,=\,b\,
{\cos\theta_0\,\sin\theta_0\,\big[b\,r_h^2\,+\,(3-2b)\,\sqrt{d^2+r_h^4\,\sin^2\theta_0}\big]\over
3(d^2\,+\,r_h^4\,\sin^2\theta_0)}\,\,r_h\,\,.
\label{theta_nh}
\eeq
To get the full $\theta(r)$ function we need to integrate numerically (\ref{eom_theta_d}) with the initial condition at $r=r_h$ given by (\ref{theta_nh}). Notice that (\ref{eom_theta_d}) depends explicitly on $r_h$ through the blackening factor $h$. It turns out that the horizon radius $r_h$ can be scaled out by an appropriate change of variables followed by a redefinition of the density $d$. Indeed, let us define the reduced variable $\hat r$ and  density $\hat d$ as:
\beq
\hat r\,=\,{r\over r_h}\,\,,
\qquad\qquad
\hat d\,=\,{d\over r_h^2}\,\,.
\label{hat_r_d}
\eeq
Then, it is readily verified that the embedding equation in terms of $\hat r$ is just (\ref{eom_theta_d}) with $r_h=1$ and $d$ substituted by $\hat d$. Other quantities can be similarly rescaled. Indeed, let us define $\hat \mu$ and $\hat m$ as:
\beq
\hat\mu\,=\,{\mu\over r_h}\,\,,
\qquad\qquad
\hat m\,=\,{m\over r_h^{b}}\,\,.
\eeq
It is straightforward to find an expression of $\hat\mu$ in terms of the rescaled quantities:
\beq
\hat \mu\,=\,{\hat d\over b}\,\int_{1}^{\infty}\,{\sqrt{b^2+\hat r^2\,h (\hat r)\,\big({d\theta\over d\hat r}\big)^{\,2}}
\over \,\sqrt{\hat d^2+\hat r^4\sin^2\theta}}
\,d\hat r\,\,.
\label{hat_chemical_pot}
\eeq
Notice also that the ratio ${\hat m}^{1\over b}/\hat \mu$ does not depend on $r_h$:
\beq
{{\hat m}^{1\over b}\over \hat \mu}\,=\,
{{ m}^{1\over b}\over  \mu}\,\,.
\eeq

\section{Zero temperature thermodynamics}
\label{zeroT_thermo}

\begin{figure}[ht]
\center
 \includegraphics[width=0.40\textwidth]{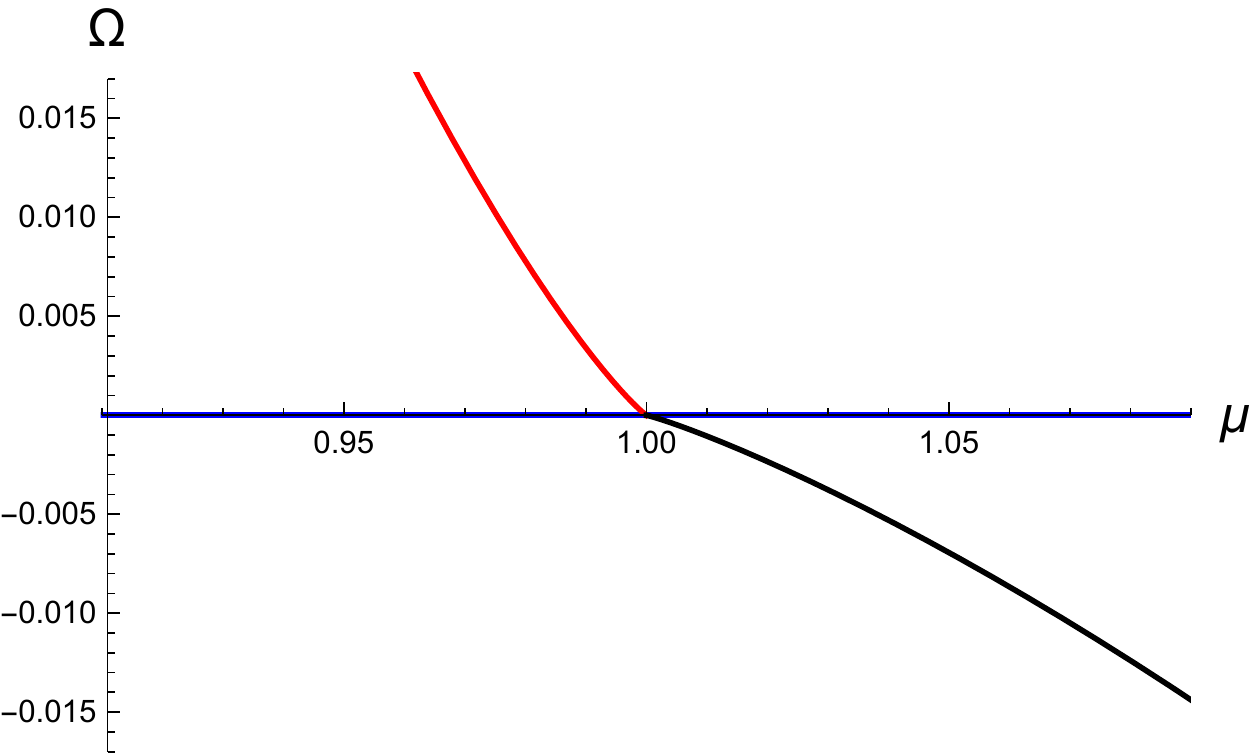}
 \qquad\qquad
  \includegraphics[width=0.40\textwidth]{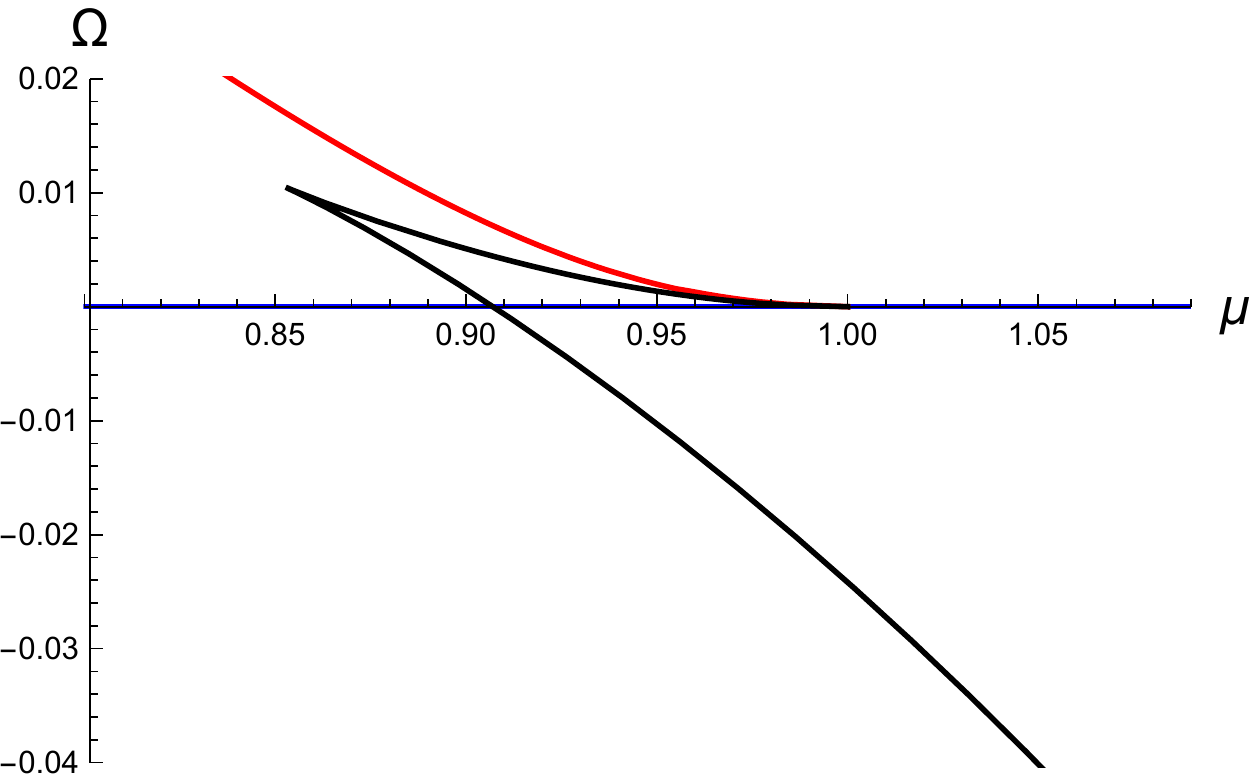}
  \caption{We plot the grand canonical potential $\Omega$ as a function of the chemical potential $\mu$ for the unflavored (left) and flavored (right) models. The black (red) curve corresponds to the black hole (brane-antibrane) embedding. The supersymmetric Minkowski embeddings $\Omega(\mu)=0$ we have represented with a blue curve on the horizontal axis. The curves for the flavored model on the right have been obtained for $b=1.25$. All curves are with $m=1$.}
\label{Omega_mu}
\end{figure}

The zero-temperature grand canonical potential $\Omega$ is given by minus the on-shell action of the probe brane:
\beq
\Omega\,=\,-S_{on-shell}\,\,.
\eeq
Notice that, as pointed out in \cite{Jokela:2012dw}, the on-shell action of our ABJM system is finite and does not need to be regulated. Indeed, the WZ term of the action serves as a regulator of the DBI term, giving rise to consistent thermodynamics. The explicit expression of $\Omega$ at zero temperature is given by:
\beq
\Omega\,=\,{\cal N}\,\int_{r_0}^{\infty}\,r^2\,\sin\theta\,\
\Bigg[{r^2\sin\theta\,\sqrt{b^2+r^2\,\theta'^2}\over \sqrt{d^2+r^4\sin^2\theta}}\,-\,
b\sin\theta\,-\,r\cos\theta\,\theta'\Bigg]dr\,\,.
\label{Omega_integral}
\eeq
Other thermodynamic properties at $T=0$ can be obtained from (\ref{Omega_integral}). For example, the pressure $P$ is just:
\beq
P\,=\,-\Omega\,\,.
\label{P_Omega}
\eeq
Moreover,  we can evaluate $\Omega$ for the different embeddings and determine the one that is favored at different values of the chemical potential. One can verify by plugging (\ref{SUSY_emb}) in (\ref{Omega_integral})  that $\Omega=0$ for the SUSY embeddings (\ref{SUSY_emb})  which have  zero density $d$ and arbitrary $\mu$. In the case of the black hole embeddings the situation varies greatly when the backreaction is included. Indeed, for the unflavored background with $b=1$ the grand canonical potential of the black hole embeddings is always negative and grows monotonically as $\mu$ decreases towards its minimal value $\mu\searrow m$, where $\Omega=0$ and $d=0$ (see Fig.~\ref{Omega_mu}, left). On the contrary, in the flavored backgrounds with $b>1$,  the grand canonical potential is negative for large values of $\mu$ and vanishes for some $\mu=\mu_c$ which corresponds to a non-zero density $d=d_c$ (see Fig.~\ref{Omega_mu}, right). From this point on, $\Omega\ge 0$, reaching a maximum  positive value, which corersponds to the minimum value of the chemical potential $\mu$. It is at this point where the black hole embedding ceases to exist as it annihilates with another (unstable) black hole embedding. This latter black hole branch is the one which connects with the Minkowski embeddings at larger $mu$, \ie\ until the grand potential reaches the value $\Omega=0$ when $\mu=m^{{1\over b}}$ and $d=0$.  The grand canonical potential for the brane-antibrane embeddings is always non-negative and decreases monotonically as $\mu$ grows ($\mu\le m^{{1\over b}}$ for these embeddings). This structure in the $(\mu,\Omega)$ plane is the well-known swallow-tail shape, typical of first-order phase transitions.

From the numerical results displayed in Fig.~\ref{Omega_mu} it is clear that the black hole embeddings are thermodynamically preferred for values of $\mu$ such that their grand canonical potential $\Omega_{bh}$ is negative. Moreover, when $\mu$ is such that $\Omega_{bh}>0$,  the Minkowski embeddings (with $d=\Omega=0$) are preferred. Notice also that the brane-antibrane configurations are always thermodynamically disfavored.  Therefore, at $\mu=\mu_c$ such that $\Omega_{bh}(\mu_c)=0$  there is a black hole-Minkowski embedding phase transition. In Fig.~\ref{Omega_mu} we see that the nature of this quantum phase transition  for the unflavored model is very different from that of the backreacted background. Indeed, in the quenched unflavored case we have a continuous second order phase transition in which the density $d$ vanishes in both phases at the transition point 
$\mu_c=m$. In section \ref{unflavored_transition} we will study in detail this quantum critical point and we will characterize the  scaling of the different physical quantities near the transition. 

In the unquenched flavored model the phase transition at $\mu=\mu_c$ is discontinuous since $d$ jumps from a non-zero value in the black hole phase to $d=0$ in the Minkowski phase. Therefore, we have a first-order phase transition, for which we will determine the latent heat and other quantities in section \ref{flavored_transition}.

\begin{figure}[ht]
\center
 \includegraphics[width=0.40\textwidth]{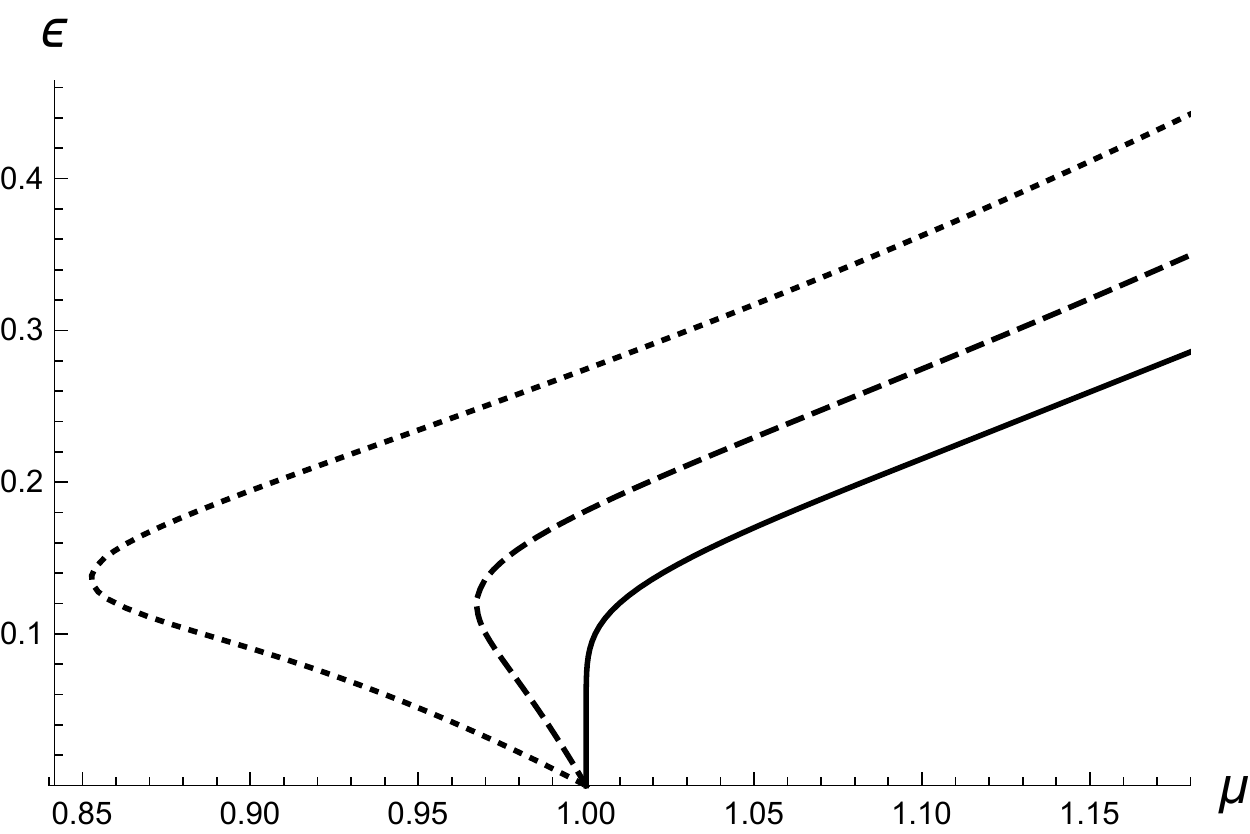}
  \caption{We depict the internal energy $\epsilon$ as a function of the chemical potential $\mu$ for the unflavored model (continuous curve) and for the flavored model with $b=1.1$ (dashed curve) and $b=1.25$ (dotted curve). In all the cases we have used $m=1$.}
\label{energy_density}
\end{figure}

Once the grand canonical potential $\Omega$ is known, we can determine other thermodynamic functions. Indeed, the charge density $\rho_{ch}$ is given by:
\beq
\rho_{ch}\,=\,-{\partial \Omega\over \partial\mu}\,\,.
\label{rho_ch_def}
\eeq
By computing numerically the derivative in (\ref{rho_ch_def}) at fixed mass $m$,  we have checked that 
$\rho_{ch}$ is related to $d$ as:
\beq
\rho_{ch}\,=\,{\cal N}\,b\,d\,\,,
\label{rho_ch_d}
\eeq
where ${\cal N}$ is the normalization constant (\ref{calN_def}).  Eq. (\ref{rho_ch_d}) confirms our identification of the constant $d$. The energy density $\epsilon$ can be obtained as:
\beq
\epsilon\,=\,\Omega\,+\,\mu\,\rho_{ch} \ .
\eeq
More explicitly, after using (\ref{Omega_integral}), (\ref{chemical_pot}),  and (\ref{rho_ch_d}), we have the following integral expression for $\epsilon$,
\beq
\epsilon\,=\,{\cal N}\,\int_{r_0}^{\infty}\,
\Big[\sqrt{b^2+r^2\,\theta'^2}\, \sqrt{d^2+r^4\sin^2\theta}\,-\,b\,r^2\,\sin^2\theta\,-\,r^3\,\sin\theta\,\cos\theta\,\theta'
\Big]dr\,\,,
\eeq
where $r_0$ is the minimal value of $r$ for the embedding. In Fig.~\ref{energy_density} we plot $\epsilon$  for black hole embeddings as a function of $\mu$, both for the quenched and unquenched model.  We notice that the energy density in the quenched theory grows monotonically with the chemical potential, starting from the value $\epsilon=0$ at the transition point at $\mu=m$. On the contrary, when dynamical quarks are added to the background, the function $\epsilon_{bh}$ is not monotonic and becomes double-valued, with a point where $\partial\epsilon/\partial\mu= \mu \partial \rho_{ch}/\partial \mu$ blows up. This is, of course, consistent with the results plotted in Fig.~\ref{plot_chemical_pot}.

\begin{figure}[ht]
\center
 \includegraphics[width=0.40\textwidth]{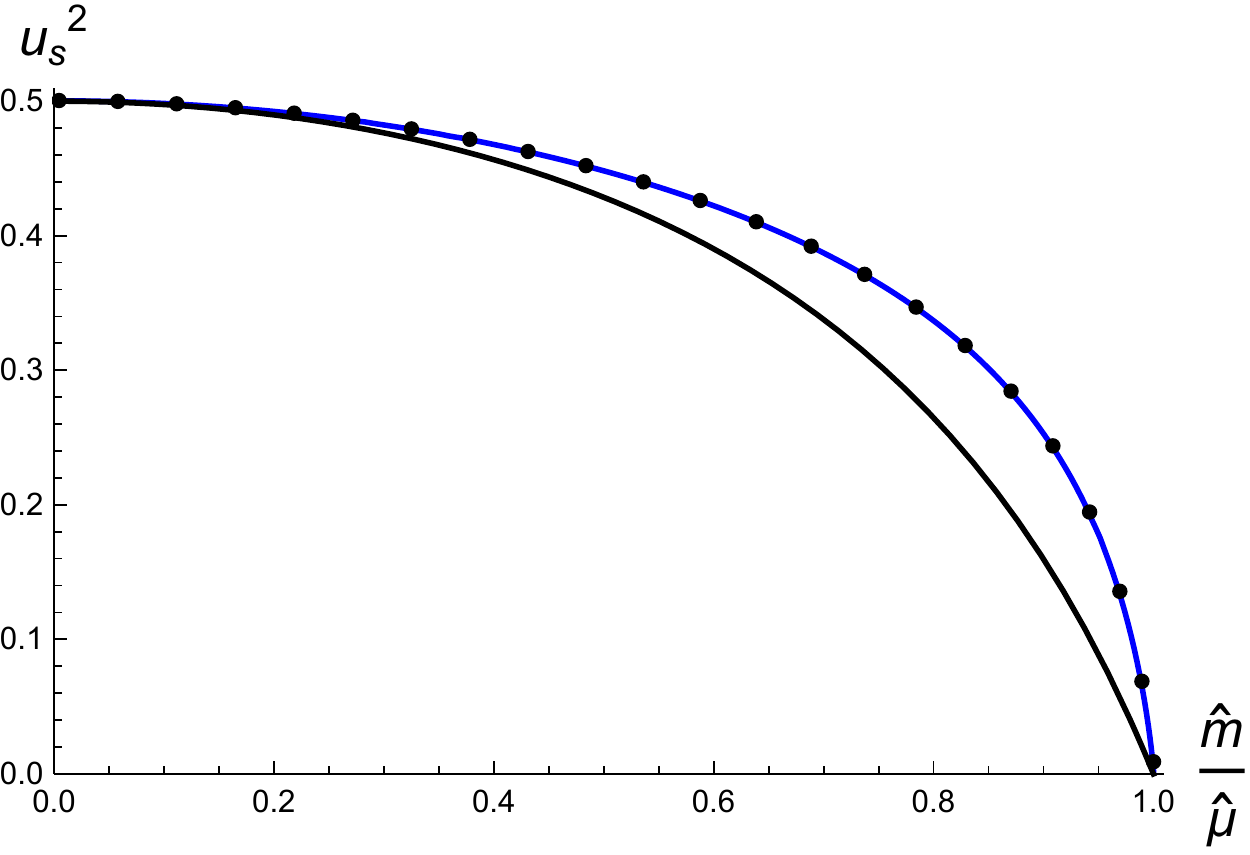}
 \qquad\qquad
  \includegraphics[width=0.40\textwidth]{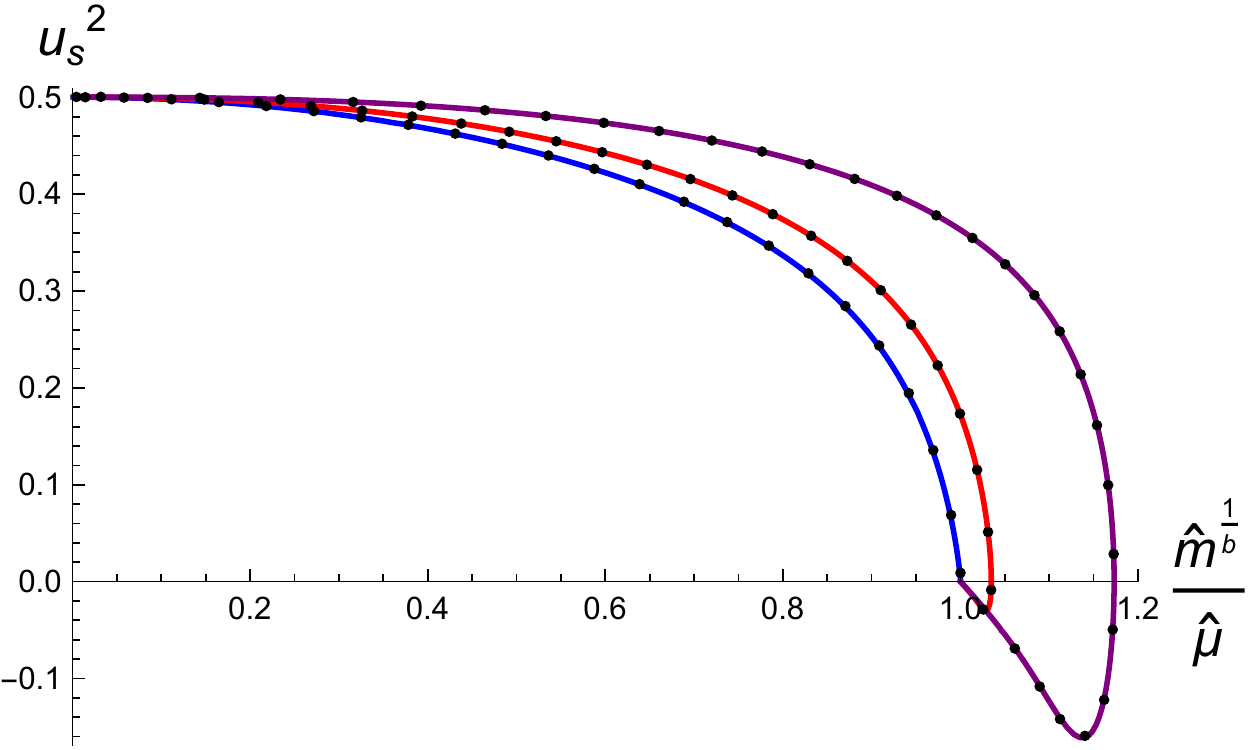}
  \caption{Left: We plot $u_s^2$ as a function of $m/\mu$ for the unflavored background (blue curve). We compare with the same quantity for the D3-D5 model (black curve).  Right: We plot $u_s^2$ for different numbers of flavors: $b=1$ (blue), $b=1.1$ (red), and $b=1.25$ (purple). In both plots the points are the values of the  square of the speed of zero sound obtained by integrating the fluctuation equations of section \ref{fluctuations}; we conclude that the speeds of first and zero sounds agree in this model.}
\label{speed_of_sound}
\end{figure}

The speed of the first sound is defined as:
\beq
u_s^2\,=\,{\partial P\over \partial \epsilon}\,\,.
\label{speed_of_first_sound}
\eeq
We evaluated   numerically the derivative in (\ref{speed_of_first_sound}) for black hole embeddings  by using 
(\ref{P_Omega}) and (\ref{Omega_integral}). The results are represented in  Fig.~
\ref{speed_of_sound}, both for the quenched and unquenched cases.  Again, they are very different in these two cases. In the quenched model $u_s^2$ is always non-negative and decreases monotonically when $m/\mu$ varies in the physical interval $[0,1]$. In Fig.~\ref{speed_of_sound} (left) we compare  $u_s^2$  for our quenched system with the corresponding values for the D3-D5 model \cite{Ammon:2012je, Itsios:2016ffv}. In the unquenched case  $u_s^2$ is not monotonic and becomes negative for small $\mu$, which again signals a discontinuous phase transition. 

\subsection{The unflavored transition}
\label{unflavored_transition}

We have shown above that the unflavored system experiences a continuous phase transition  at $\mu=m$ and $T=0$. In this section we look in more detail the behavior of the system near this quantum critical point. Accordingly, let us define $\bar\mu$ as:
\beq
\bar\mu\,=\,\mu\,-\,m\,\,.
\label{mu_bar_def}
\eeq
Clearly $\bar\mu=0$ is the location of the phase transition. Therefore,  we expect that the grand canonical potential $\Omega$  behaves in a non-analytic form  near $\bar\mu=0$. We assume that the system displays a scaling behavior near the critical point. The goal of this section is to characterize this behavior  in terms of a set of critical exponents. 

Let us consider a system with hyperscaling violation exponent $\theta$ and dynamical exponent $z$ in $n$ spatial dimensions ($n=2$ in our case). Recall that in such a system $n-\theta$ is the effective number of spatial dimensions near the critical point and $z$ is the effective dimension of the energy.  Therefore $[\bar\mu]=z$ and the energy densities (such as our grand canonical potential $\Omega$) should have a dimension equal to $n-\theta+z$. These dimension assignments allow us to write $\Omega$ near $\bar\mu=0$ as:
\beq
\Omega\,\approx\, -C\,\bar\mu^{{n+z-\theta\over z}}\,\Big(\big|\log {\bar\mu\over m}\big|\Big)^{-\zeta}\,\,,
\label{Critical_Omega}
\eeq
where $C>0$ is a constant.  Eq. (\ref{Critical_Omega})  is a generalization of the expression written in \cite{Ammon:2012je}  by including a logarithmic multiplicative term with some new exponent $\zeta$. We show below that $\zeta$ cannot be zero in our ABJM case.  This is to be compared with the D$p$-D$q$ systems studied in \cite{Ammon:2012je, Itsios:2016ffv}, where  $\zeta=0$. Similar multiplicative logarithmic corrections to the scaling has been studied in general in \cite{Kenna} for thermal phase transitions. 

\begin{figure}[ht]
\center
 \includegraphics[width=0.40\textwidth]{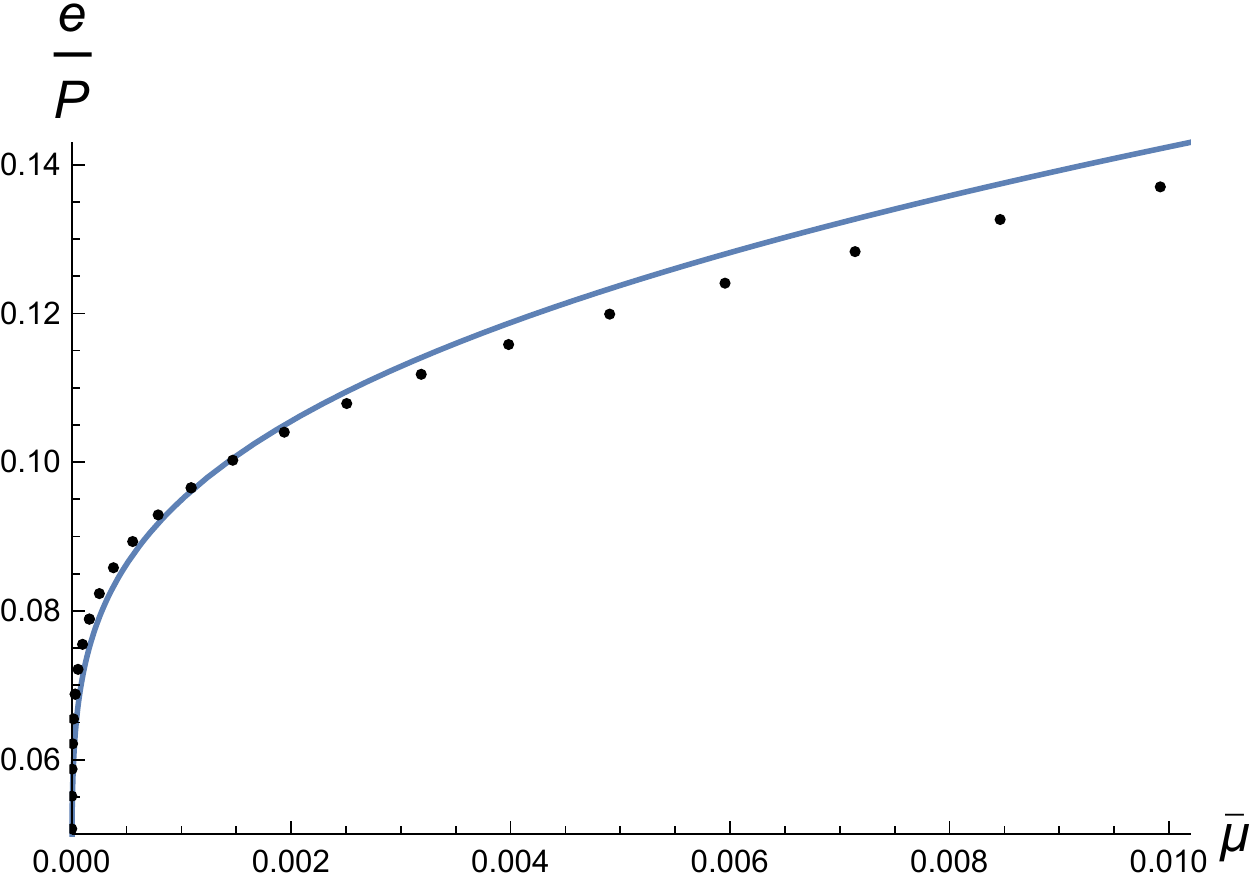}
 \qquad\qquad
  \includegraphics[width=0.40\textwidth]{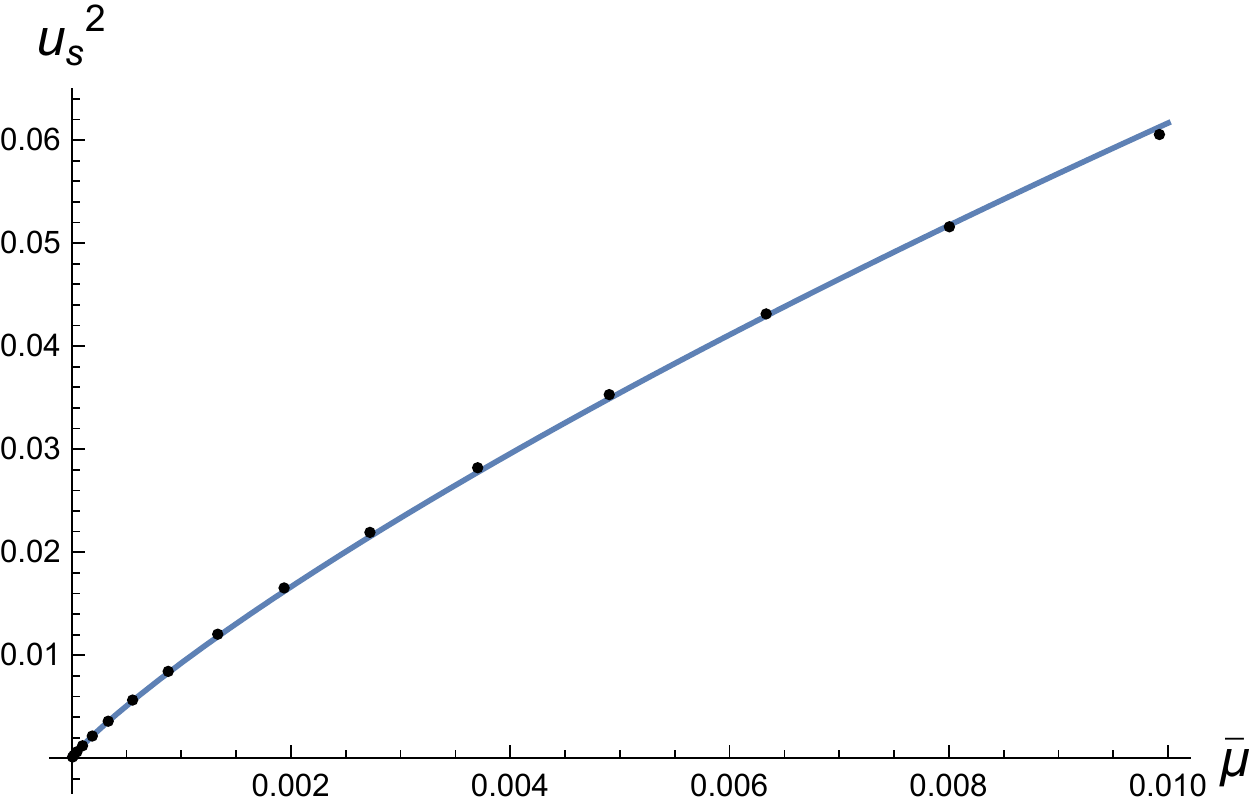}
  \caption{On the left we plot the numerical values of the  ratio $e/P$ as a function of $\bar\mu$. The continuous line 
  is a fit to the expression written in (\ref{rho_e_over_P}). The value of $\zeta$ obtained in this fit is $\zeta=0.65575$.  On the right we plot the values of $u_s^2$, together with the scaling expression (\ref{us_scaling}) for  $\zeta=0.74689$.
  }
\label{e/p_cs}
\end{figure}

The charge density $\rho_{ch}={\cal N}\,d$  is obtained by computing the derivative of $\Omega$ with respect to $\bar\mu$. We get:
\beq
{\cal N}\,d\,\approx\,C\,\bar\mu^{{n-\theta\over z}}\,
\,\Big(\big|\log {\bar\mu\over m}\big|\Big)^{-\zeta}\,
\Big[1+{n-\theta\over z}\,+\,{\zeta\over \big|\log {\bar\mu\over m}\big|}\Big]\,\,.
\label{critical_d}
\eeq
Let us  next consider, following \cite{Ammon:2012je}, the non-relativistic energy density $e$,  defined  as:
\beq
e\,=\,\epsilon\,-\,\rho_{ch}\,m\,=\,\Omega+\,\rho_{ch}\,\bar\mu\,\,.
\eeq
Near the critical point, $e$ behaves as:
\beq
e\,\approx\, 
C\,\bar\mu^{{n+z-\theta\over z}}\,\Big(\big|\log {\bar\mu\over m}\big|\Big)^{-\zeta}\,
\Big[
{n-\theta\over z}\,+\,{\zeta\over  \big|\log {\bar\mu\over m}\big|}
\Big]\,\,,
\eeq
and it is very convenient to consider the ratio $e/P$, which is given by:
\beq
{e\over P}\,\approx\,{n-\theta\over z}\,+\,{\zeta\over  \big|\log {\bar\mu\over m}\big|}\,\,.
\eeq
If $\theta\not=n$ the ratio $e/P$ reaches a constant non-vanishing value as $\bar\mu\to0$. This is clearly not the case  for our system, as illustrated in Fig.~\ref{e/p_cs}. Therefore,  our system should have $\theta=2$. Moreover, the  logarithmic exponent $\zeta$ should be non-zero and positive.\footnote{Indeed, if we had $\theta=2$ and $\zeta=0$ the charge density $d$ in (\ref{critical_d})  would be non-zero at the critical point, which is not the case for our ABJM system.} Therefore we get the following leading behavior for our system:
\beq
\rho_{ch}\,=\,{\cal N}\,d\,\approx \,{C\over \Big(\big|\log {\bar\mu\over m}\big|\Big)^{\zeta}}\,\,,
\qquad\qquad\qquad\qquad
{e\over P}\,\approx {\zeta\over \big|\log {\bar\mu\over m}\big|}\,\,.
\label{rho_e_over_P}
\eeq
We can also compute the speed of sound $u_s$ near the critical point by using (\ref{speed_of_first_sound}), with the result:
\beq
u_s^2\,\approx\, {1\over \zeta}\,{\bar\mu \big|\log {\bar\mu\over m}\big|\over m+\bar\mu}\,
\Big[\,1-{1\over 1+\zeta+\big|\log {\bar\mu\over m}\big|}\,\Big]\,\,,
\eeq
which, at leading order for $\bar\mu\to 0$ becomes simply:
\beq
u_s^2\,\approx\, {1\over \zeta}\,{\bar\mu\over m}\,\big|\log {\bar\mu\over m}\big|\,\,.
\label{us_scaling}
\eeq

To determine the value of the exponent $\zeta$ we can fit the numerical values of $e/P$ and $u_s^2$ near $\bar\mu=0$ to our scaling expressions (\ref{rho_e_over_P}) and (\ref{us_scaling}). Due to the logarithmic behavior of these quantities we must explore very small values of $\bar\mu$. The results of these fits is shown in Fig.~\ref{e/p_cs}. The values of $\zeta$ obtained are in the range $\zeta=0.65-0.75$.

Let us determine, following the reasoning in \cite{Ammon:2012je}, the dynamical critical exponent $z$ by dimensional analysis of   the dispersion relation of the sound mode, which is of the form $\omega\,=\,u_s\, k$, where 
$u_s$ is given by (\ref{us_scaling}) near the critical point $\bar\mu=0$. Actually, we will see that the speed of the zero sound, obtained by numerical integration of the fluctuation equations of the probe brane, is exactly the same as the one determined by (\ref{speed_of_first_sound}). Near $\bar\mu=0$ eq. (\ref{us_scaling}) tells us that $u_s\sim \sqrt{\bar\mu}$  (times a logarithmic correction) and, since $[\omega]=[\bar\mu]=z$ and $[k]=1$,  the dimensional consistency of the dispersion relation $\omega\,=\,u_s\, k$ implies that $z=2$. Therefore, the values of $\theta$ and $z$ for our system are:
\beq
\theta=2\,\,,
\qquad\qquad
z\,=\,2\,\,.
\label{theta_z}
\eeq
Notice that the value of $\theta$ just found differs from the value   $\theta=1$ obtained in \cite{Ammon:2012je} for the conformal systems D3-D7 and D3-D5.

Let us now consider the system  at small non-zero temperature $T\ll \bar \mu$. We can evaluate the free energy 
$f$ at first order in $T$ by using the results of \cite{Karch:2009eb}. Notice that at $T=0$, $f=\epsilon$. Indeed, according to the analysis of
\cite{Karch:2009eb}, when $T$ is small the free energy density can be approximated as:
\beq
f(\mu,m,T)\,=\,f(\mu,m,T=0)\,+\,\pi\, \rho_{ch}\,T\,+\,{\mathcal O}(T^2)\,\,.
\eeq
Then, the non-relativistic free energy  density is given by:
\beq
f_{non-rel}(\mu,m,T)\,=\,f(\mu,m, T)\,-\, \rho_{ch}\,m\,=\,e\,+\,
\pi\, \rho_{ch}\,T\,+\,{\mathcal O}(T^2)\,\,.
\label{f_non_rel_def}
\eeq
Evaluating  the right-hand side of  (\ref{f_non_rel_def}) for our system, we get the following expression of $f_{non-rel}$ for small 
$\bar\mu$ and $T/\bar\mu$:
\beq
f_{non-rel}(\mu,m,T) = C\,{\bar\mu\over 
\Big(\big|\log {\bar\mu\over m}\big|\Big)^{\zeta+1}}\,\Big[\zeta\,+\,
\pi\,\big|\log {\bar\mu\over m}\big|\,{T\over \bar\mu}\,+\,\cdots\Big]\,\,.
\label{f_non_rel_ABJM}
\eeq
On general grounds, near a quantum phase transition the free energy density should behave as a homogeneous function when the control parameter $\bar\mu$ and the temperature $T$ are scaled as $\bar\mu\to \Lambda^{{1\over \nu}}\,\bar\mu$,  $T\to \Lambda^{z}\, T$, where $\nu$ is the critical exponent that characterizes the divergence of the correlation length $\xi\sim (T-0)^{-\nu}$ \cite{Vojta}.  Eq. (\ref{f_non_rel_ABJM}) is the first order term of a Taylor expansion of the scaling function of $f_{non-rel}$. If we disregard the logarithmic terms in (\ref{f_non_rel_ABJM}) (which give rise to  subleading terms when $\bar\mu\to 0$), it follows that $T$ and $\bar\mu$ should be scaled by the same power of the scale factor $\Lambda$. Since $z=2$ for our system, we must have $\nu=1/2$. Eq.  (\ref{f_non_rel_ABJM}) also determines the value of the exponent $\alpha$ which characterizes the scaling of the heat capacity $c_V\sim (T-0)^{-\alpha}$. Indeed, according to the analysis of \cite{Ammon:2012je} the global power of $\bar\mu$  in $f_{non-rel}(\mu,m,T)$  should be $2-\alpha$. If we ignore again the logarithmic correction, this prescription gives $\alpha=1$. Therefore, we have obtained  that the critical exponents $\alpha$ and 
$\nu$ are given by
\beq
\alpha\,=\,1\,\,,
\qquad\qquad
\nu\,=\,{1\over 2}\,\,.
\label{alpha_nu}
\eeq
Notice that the values of $\theta$, $z$, $\alpha$, and $\nu$ listed in (\ref{theta_z})  and (\ref{alpha_nu}) satisfy the hyperscaling relation 
\beq
(n+z-\theta)\,\nu\,=\,2-\alpha\,\,,
\eeq
with $n=2$.

\subsection{The flavored transition}
\label{flavored_transition}

We already pointed out above that the black hole-Minkowski phase transition with dynamical quarks in the background is of first order. At the transition point  the density jumps from being $d=d_c\not=0$ in the black hole phase to $d=0$ in the Minkowski phase. We have investigated numerically the dependence of $d_c$ on $\hat\epsilon$ and $m$ and we found that, with big accuracy, this dependence can be written as:
\beq
d_c(\hat\epsilon, m)\,=\,\tilde d_{c}(\hat \epsilon)\,\,
m^{{2\over b}}\,=\,\tilde d_{c}(\hat \epsilon)\,\,m_q^2\,\,,
\label{quadratic_law}
\eeq
where $m_q=m^{{1\over b}}$ is proportional to the physical mass of the quarks. Notice that the dependence on $m$ written in (\ref{quadratic_law}) is the one expected by the rescaling argument given at the end of section \ref{zeroT_embeddings}.

\begin{figure}[ht]
\center
 \includegraphics[width=0.40\textwidth]{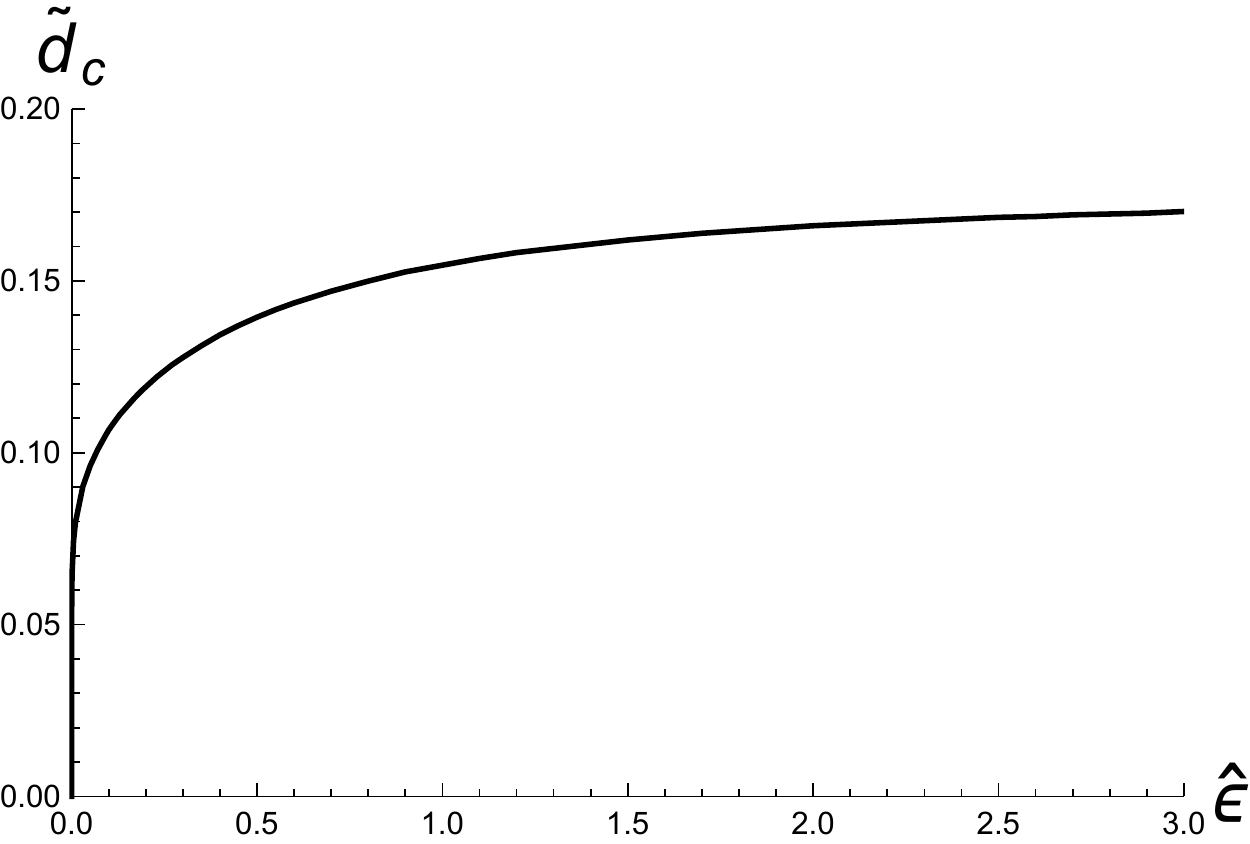}
 \qquad\qquad
  \includegraphics[width=0.40\textwidth]{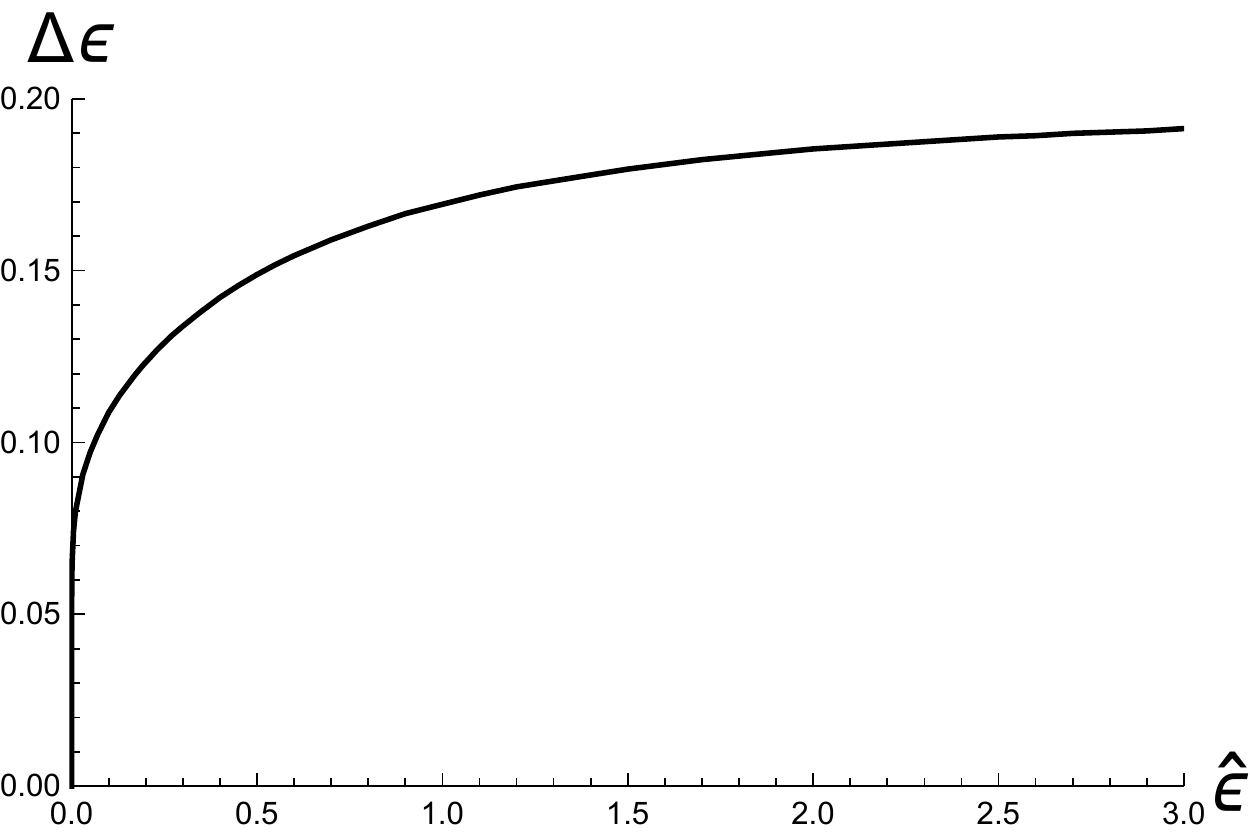}
  \caption{Left: We plot the function $\tilde d_{c}(\hat \epsilon)$ introduced in (\ref{quadratic_law}) at $m=1$. Right: We depict the latent heat $\Delta \epsilon$ for $m=1$ as a function of $\hat\epsilon$.}
\label{flavored_transition_plots}
\end{figure}

The flavor dependent coefficient of the quadratic law (\ref{quadratic_law}) grows monotonically  with $\hat\epsilon$, as shown in Fig.~\ref{flavored_transition_plots} (left).  For small $\hat\epsilon$ this growth is very fast and saturates very quickly for larger values of the deformation parameter.  

The phase transition occurs at a critical  chemical potential $\mu_c< m^{{1\over b}}=m_q$. Actually, 
the ratio $\mu_c/m_q$ is a decreasing function of $\hat \epsilon$ which approaches the value $\mu_c/m_q\approx 0.9$ when $\hat\epsilon\to\infty$. It is also interesting to point out that the value of $\mu$ where the speed of sound 
vanishes (see Fig.~\ref{speed_of_sound})) corresponds to the turning point of $\Omega$ as a function of $\mu$ for a black hole embedding, \ie\ to the minimum value of $\mu$ for such embeddings. The phase transition occurs for  a value of $\mu$ close to its lowest value where $u_s^2$ is still positive. Moreover, it follows from the above discussion that $\mu_c\sim d_c^{{1\over 2}}$.

We also studied the latent heat of the phase transition, \ie\ the difference $\Delta\epsilon$ of the internal energy of the two phases. Notice that, as $\Omega=0$ in both phases at the transition point and $\rho_{ch}=0$ in the Minkowski phase,  $\Delta\epsilon$ is simply obtained by evaluating $\mu\,\rho_{ch}$ at the black hole side of the transition:
\beq
\Delta \epsilon\,=\,(\mu\,\rho_{ch})_{bh}\,\,.
\eeq
The behavior of this quantity with the number of flavors  when $m=1$ is displayed in Fig.~\ref{flavored_transition_plots} (right).  We notice that the latent heat resembles the behavior of the critical density. 
We have also verified that $\Delta \epsilon$ grows with the quark mass as $\Delta \epsilon\sim m_q^3\,=\,m^{{3\over b}}$.

Most of the figures that we have presented above have been produced using $m=1$. It is however simple to obtain the results for any value of $m$ by using the rescaling argument presented above. Indeed, one can readily show that the different quantities scale with $m_q=m^{1\over b}$ as:
\be
 \epsilon\sim \Omega \sim m_q^3 \ , \ \mu \sim m_q \ , \ d\sim m_q^2 \ .
\ee
We have checked that this behavior is confirmed by our numerical results.

\section{Charge susceptibility and diffusion constant}
\label{diffusion}

Let us now consider the system at non-zero temperature and compute the charge susceptibility, which is defined as:
\beq
\chi\,=\,{\partial\rho_{ch}\over \partial\mu}\,\,.
\eeq
Taking into account that the charge density $\rho_{ch}$ is related to $d$ as $\rho_{ch}\,=\,{\cal N}\,b\,d$ (\ref{rho_ch_d}), we can rewrite this last expression as:
\beq
\chi^{-1}\,=\,{1\over {\cal N}\,b}\,{\partial \mu\over \partial d}\,\,.
\label{charge_sus_def}
\eeq
We now evaluate  explicitly the derivative in (\ref{charge_sus_def}) as:
\beq
{\partial \mu\over \partial d}\,=\,\int_{r_h}^{\infty}\,
{\partial  A_t'\over \partial d} \,dr\,\,.
\label{d_mu_d}
\eeq
The derivative inside the  integral in (\ref{d_mu_d})  can be computed directly. We get:
\beq
{\partial  A_t'\over \partial d}\,=\,{\sqrt{\Delta}\over b}\,
{r^2\sin\theta\over d^2+r^4\sin^2\theta}\,
\Bigg[1\,-\,d\Bigg(\cot\theta\,{\partial\theta\over \partial d}\,-\,{r^2\,h\,\theta'\over \Delta}\,
{\partial\theta'\over \partial d}\Bigg)\Bigg]\,\,.
\eeq
where $\Delta$ is defined as:
\beq
\Delta\,\equiv\,b^2(1-A_t'^{\,2})\,+\,r^2\,h\,\theta'^{\,2}\,=\,
{r^4\sin^2\theta(b^2+r^2\,h\,\theta'^{\,2})\over
d^2+r^4\sin^2\theta}\,\,.
\label{Delta_definition}
\eeq
Thus, the charge susceptibility can be written in the form:
\beq
\chi^{-1}\,=\,{1\over {\cal N}}\,\int_{r_h}^{\infty}\,dr
{\sqrt{\Delta}\over b^2}\,
{r^2\sin\theta\over d^2+r^4\sin^2\theta}\,
\Bigg[1\,-\,d\Bigg(\cot\theta\,{\partial\theta\over \partial d}\,-\,{r^2\,h\,\theta'\over \Delta}\,
{\partial\theta'\over \partial d}\Bigg)\Bigg]\,\,.
\label{susceptibility}
\eeq

\begin{figure}[ht]
\center
 \includegraphics[width=0.40\textwidth]{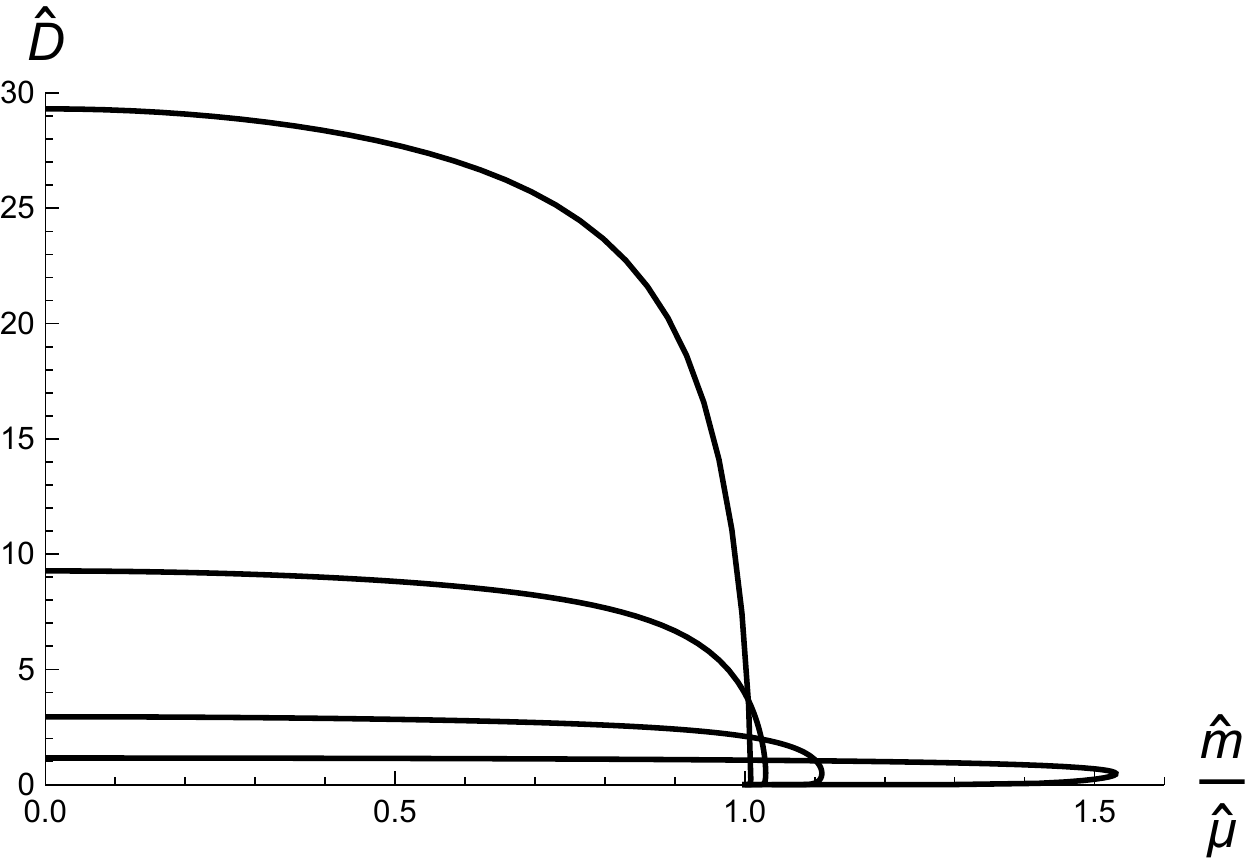}
 \qquad\qquad
  \includegraphics[width=0.40\textwidth]{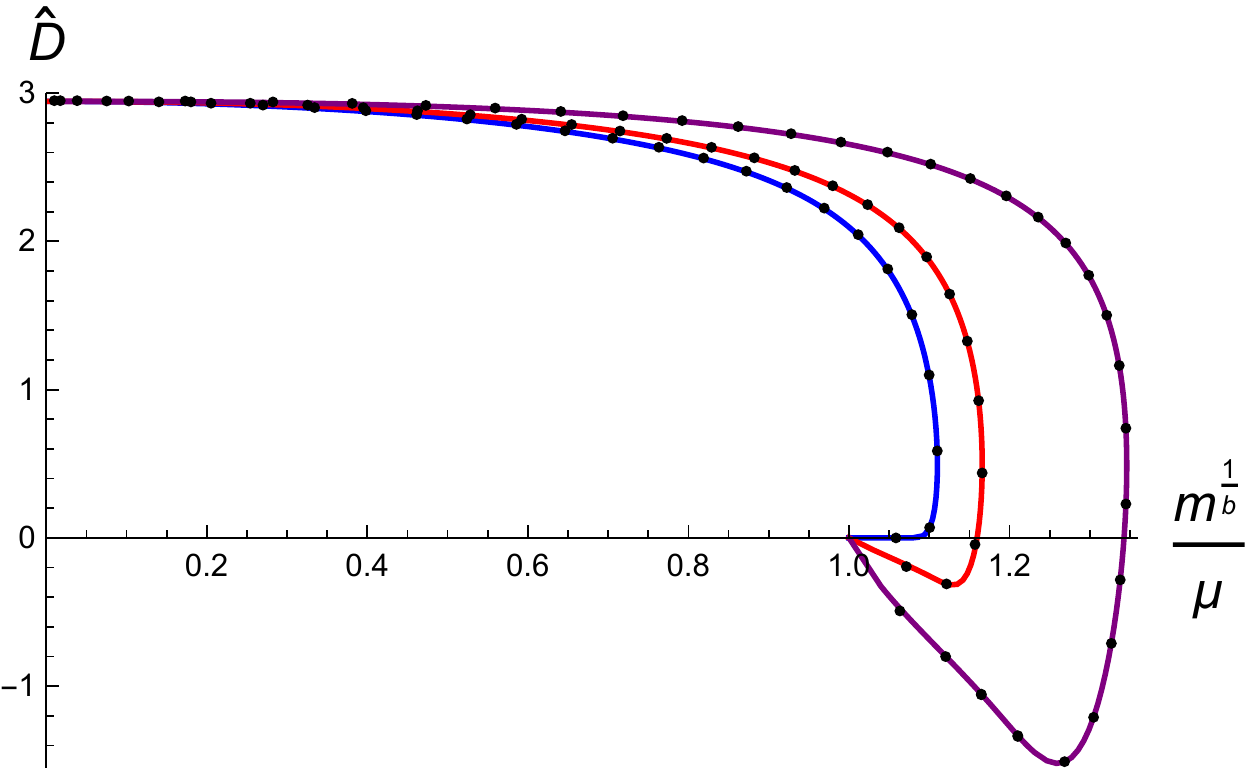}
  \caption{We plot the rescaled diffusion constant $\hat D=r_h\,D$ as a function of the ratio of the mass and the chemical potential. Left: We plot the values of $\hat D$ for the unflavored theory for different values of the rescaled density $\hat d=d/r_h^2$. The values of $\hat d$ in this plot  are $\hat d=1,10,100,1000$ (bottom-up). Right: We plot the values of $\hat D$ for $\hat d=10$ and for different number of flavors: $b=1$ (blue), $b=1.1$ (red), and $b=1.25$ (purple), inside-out. The continuous curves are obtained by the Einstein relation and the points correspond to the diffusive fluctuation modes of the probe.}
\label{Diffusion_constant_plots}
\end{figure}

The charge diffusion constant $D$ can be related to the charge susceptibility and to the DC conductivity $\sigma$ by the Einstein relation:
\beq
D\,=\,\sigma\,\chi^{-1}\,\,.
\label{Einstein_rel}
\eeq
The value of $\sigma$ can be obtained from the two-point correlators  of the transverse currents. This calculation is performed in detail in appendix \ref{fluctuation_apendix}. Alternatively, $\sigma$ can be computed by employing the Karch-O'Bannon method \cite{Karch:2007pd}, as was done for the ABJM model in \cite{Bea:2014yda}.  The result obtained by these two methods agree and is given by:
\beq
\sigma\,=\,{\cal N}\,{b\over r_h^2}\,\sqrt{d^2+r_h^4\,\sin^2\theta_h}\,\,.
\label{dc_conductivity}
\eeq
We can now plug  (\ref{susceptibility}) and (\ref{dc_conductivity}) into the right-hand side of (\ref{Einstein_rel}) to get the diffusion constant $D$. The final result is:
\beq
D\,=\,{\sqrt{d^2+r_h^4\,\sin^2\theta_h}\over b\,r_h^2}\,
\int_{r_h}^{\infty}\,dr
\,{r^2\sin\theta\sqrt{\Delta}\over d^2+r^4\sin^2\theta}\,
\Bigg[1\,-\,d\Bigg(\cot\theta\,{\partial\theta\over \partial d}\,-\,{r^2\,h\,\theta'\over \Delta}\,
{\partial\theta'\over \partial d}\Bigg)\Bigg]\,\,.
\label{Diffusion_Einstein}
\eeq
In the case of massless quarks, the embedding is just $\theta={\rm constant}=\pi/2$ and the integral (\ref{Diffusion_Einstein}) can be evaluated in analytic form. We get:
\beq
D_{m=0}\,=\,{\sqrt{d^2+r_h^4}\over r_h^3}\,
F\Big({3\over 2}, {1\over 2}; {5\over 4};-{d^2\over r_h^4}\Big)\,\,.
\label{massless_diffusion_constant}
\eeq
In the general case of massive quarks we have evaluated (\ref{Diffusion_Einstein}) numerically for the unflavored and  flavored backgrounds as a function of the chemical potential. The results of these calculations are displayed in Fig.~\ref{Diffusion_constant_plots}. In the unflavored background $D$ is always non-negative and vanishes when 
$\mu=m$ (see Fig.~\ref{Diffusion_constant_plots}, left).  On the contrary, when $N_f\not= 0$ the diffusion constant  
is maximal for large chemical potential (and given by the massless value (\ref{massless_diffusion_constant})) and 
becomes negative after $\mu$ reaches its minimal value, which means that the system becomes unstable and that the first-order phase transition at $T=0$ survives at non-zero temperature. In the next section  we obtain the diffusion constant by looking at the fluctuation modes of the probe in the hydrodynamical regime. The corresponding values of $D$ are also plotted in Fig.~\ref{Diffusion_constant_plots}, where we notice that they agree perfectly with the values found above by using Einstein relation.

\section{Fluctuations}
\label{fluctuations}

We now want to carry out a dynamic (\ie\ time-dependent) study of our system,  to complement the static analysis performed so far. Accordingly, let us consider the generic $T\not=0$ background and let us allow the probe brane to fluctuate around the black hole embeddings described in section \ref{Probes}. In general, the equations of motion of these fluctuations are very complicated since the different fluctuation modes are coupled. However, there are certain modes that can be decoupled from the rest and, therefore, they constitute a consistent truncation of the general system of equations. In this section we will study one of these restricted sets of fluctuations, which involves the gauge   field  $A$ and the transverse scalar $\theta$.  These fields take the form:
\beq
A\,=\,L^2\,\big[A_t(r)\,dt\,+\,a_t(t,x,r)\,dt+a_x(t,x,r)\,dx+a_r(t,x,r)\,dr]\,\,,
\quad
\theta\,=\,\theta(r)+\lambda(t,x,r)\,\,\,,
\label{longitudinal_fluct_ansatz}
\eeq
where $a_t$, $a_x$, $a_r$, and $\lambda$ are the first-order perturbations. One can check that the ansatz (\ref{longitudinal_fluct_ansatz}) is indeed a consistent truncation of the equations of motion. These truncated equations  can be derived from a second order Lagrangian density ${\cal L}^{(2)}$, which is derived in detail in appendix \ref{fluctuation_apendix}.  The expression for  ${\cal L}^{(2)}$ is:
\bear
&&{\cal L}^{(2)}=-{\cal N}\,r^2\,\sin\theta\sqrt{\Delta}\Bigg[{1\over 4}\,{\cal G}^{nm}\,{\cal G}^{pq}\,f_{mq}\,f_{np}\,+\,{L^2\over 2b^2}\,
\Big(1\,-\,{r^2\,h\,\theta'^{\,2}\over \Delta}\Big)\,{\cal G}^{mn}
\partial_m\lambda\,\partial_n\lambda \rc
&&
\qquad\qquad\qquad\qquad
+\Big(\big(b-{3\over 2}\big)\,{\sin\theta\over \sqrt{\Delta}}\,+\,
{r^2\,h\,\theta'^{\,2}-\Delta\over 2\sin^2\theta\Delta}\Big)\lambda^2- 
{d^2\theta'^{\,2}\over 2\, b^2\, r^4\sin^2\theta\Delta}\,(\partial_t\lambda)^2
\rc
&&
\qquad\qquad\qquad\qquad
+{d\theta'\over b r^2\sin\theta\sqrt{\Delta}}\,{\cal G}^{mn}\partial_m \lambda\, f_{nt}
+{b\,d\cot\theta\over L^2 r^2\sin\theta\sqrt{\Delta}}\,\lambda f_{tr}\Bigg]\,\,,
\label{total_lag_fuct}
\eear
where ${\cal G}^{mn}$ is the open string metric defined in (\ref{open_string_metric_def}), $f_{mn}$ is the field strength for $a_m$ ($f=L^2 da$)  and $\Delta$ is given by (\ref{Delta_definition}). Let us now write the different equations of motion which can be derived from the total Lagrangian (\ref{total_lag_fuct}). The non-zero values of ${\cal G}^{mn}$ are written in (\ref{open_string_metric_components}).  First of all, we write the equation of motion for $a_r$ (in the $a_r=0$ gauge):
\beq
{b^2+r^2\,h\,\theta'^{\,2}\over h\,\Delta}\,\partial_t\,a_t'\,-\,\partial_x\,a_x'\,-\,
{d\over b\sin\theta\sqrt{\Delta}}\Big(\theta'\,\partial_t\,\lambda'\,-\,{\Delta\over r^2\,h}\,\cot\theta\,
\partial_t\lambda\Big)\,=\,0\,\,.
\label{eom_ar}
\eeq
The equation of motion for $a_t$ is:
\bear
&&\partial_r\,\Bigg[
{b^2\,r^2\,\sin\theta\over \Delta^{{3\over 2}}}\,(b^2+r^2\,h\,\theta'^2)\,a_t'\,
+\,d\,b\Big(\cot\theta\,\lambda\,-\,{r^2\,h\,\theta'\over \Delta}\,\lambda'\Big)\Bigg]  \rc
&&
\qquad\qquad
+{\sin\theta(b^2+r^2\,h\,\theta'^{\,2})\over r^2\, h\, \sqrt{\Delta}}\,
\partial_x(\partial_x a_t\,-\,\partial_t a_x)\,-\,
{d\,\theta'\over b\,r^2}\,\partial_x^2\,\lambda\,=\,0\,\,, 
\label{eom_at}
\eear
while the equation of motion for $a_x$ becomes:
\beq
\partial_r\,\Big({b^2\,r^2\,h\,\sin\theta\over \sqrt{\Delta}}\,a_x'\Big)\,+\,
{\sin\theta(b^2+r^2\,h\,\theta'^2)\over r^2 \,h\,\sqrt{\Delta}}\,
\partial_t(\partial_x\,a_t-\partial_t a_x)\,-\,{d\,\theta'\over b\,r^2}\,\partial_t\partial_x\lambda\,=\,0\,\,.
\label{eom_ax}
\eeq
Finally, the equation of motion for the scalar $\lambda$ is:
\bear
&&\partial_r\,\Bigg[{r^2\sin\theta h\over \sqrt{\Delta}}\,\Big[r^2\,
\Big(1\,-\,{r^2\,h\,\theta'^{\,2}\over \Delta}\Big)\lambda'+
{b\,d\,\theta'\over \sin\theta\sqrt{\Delta}}\,a_t'\Big]\Bigg] \\
&&
\qquad
+b\,d\cot\theta\,a_t'\,
+{d\theta'\over b\,r^2}\,\partial_x(\partial_x a_t-\partial_t a_x) +r^2\sin\theta\,\sqrt{\Delta}\,\Big[\big(3-2b){\sin\theta\over \sqrt{\Delta}}\,+\,
{\Delta-r^2\, h\,\theta'^2\over \sin^2\theta\Delta}\,\Big]\,\lambda \rc
&&
+ {\sin\theta\over b^2\sqrt{\Delta}}\Big[
{(b^2+r^2\,h\,\theta'^2)(r^2\,h\,\theta'^{\,2}-\Delta)\over h\,\Delta}\,-\,
{d^2\,\theta'^{\,2}\over r^2\sin^2\theta}\,\Big]\partial_t^2\lambda+
{\sin\theta\sqrt{\Delta}\over b^2}\,
\Big(1\,-\,{r^2\,h\,\theta'^{\,2}\over \Delta}\Big)\partial_x^2\lambda\,=\,0\,\,.\nonumber
\label{eom_lambda}
\eear
Let us  next Fourier transform the gauge field  and the scalar to momentum space as:
\bear
a_\nu(r, t, x) & = & \int {d\omega\,dk\over (2\pi)^2}\,
a_\nu(r, \omega, k)\,e^{-i\omega\,t\,+\,i k x}  \rc
\lambda(r, t, x) & = & \int {d\omega\,dk\over (2\pi)^2}\,
\lambda(r, \omega, k)\,e^{-i\omega\,t\,+\,i k x}\,\,,
\eear
and  define the electric field $E$ as the following  gauge-invariant combination:
\beq
E\,=\,k\,a_t\,+\,\omega\,a_x\,\,.
\label{E_at_ax}
\eeq
Then, the  equation of motion for $a_r$ in momentum space is:
\beq
{b^2+r^2\,h\,\theta'^{\,2}\over h\,\Delta}\,\omega\,a_t'\,+\,k\,a_x'\,-\,
{\omega\,d\over b\sin\theta\sqrt{\Delta}}\Big(\theta'\,\lambda'\,-\,{\Delta\over r^2\,h}\,\cot\theta\,
\lambda\Big)\,=\,0\,\,.
\eeq
We now combine this last equation with the definition of $E$. We get $a_t'$ and $a_x'$ as functions of $E$ and $\lambda$:
\bear
a_t' & = & {k\,h\,\Delta\over \Delta h k^2-(b^2+r^2 h\theta'^2)\omega^2}\,E'-
{\omega^2\,h\,d\,\sqrt{\Delta}\over b\big[\Delta h k^2-(b^2+r^2 h \theta'^2)\omega^2\big]\sin\theta}\,
\big(\theta'\lambda'-{\Delta\over r^2 h}\cot\theta\lambda\big) \rc
a_x' & = & {-(b^2+r^2 h \theta'^2)\omega\over \Delta h k^2-(b^2+r^2 h \theta'^2)\omega^2}\,E'+{\omega\,k\,h\,d\,\sqrt{\Delta}\over b\big[\Delta h k^2-(b^2+r^2 h \theta'^2)\omega^2\big]\sin\theta}\big(\theta'\lambda'-{\Delta\over r^2h}\cot\theta\lambda\big) \ .\rc
\label{at_ax_E}
\eear
After using these equations, it is easy to check that (\ref{eom_at}) and (\ref{eom_ax}) become equivalent and equal to the following differential equation for the electric field $E$:
\bear
&&
\partial_r\Bigg[{b^2\,r^2\,h\over 
(b^2+r^2\,h\,\theta'^{\,2})\omega^2-\Delta\,h\,k^2}\Bigg({\sin\theta\over \sqrt{\Delta}} (b^2+r^2\,h\,\theta'^2)\,E'\,-\,{k\,d\,h\over b}\,\big(\theta'\,\lambda'-{\Delta\over r^2\,h}\cot\theta\lambda\big)
\Bigg)\Bigg] \rc
&&
\qquad\qquad\qquad\qquad\qquad\qquad\qquad\qquad
+{\sin\theta (b^2+r^2\,h\,\theta'^{\,2})\over r^2\,h\,\sqrt{\Delta}}\,E\,-\,
{d\theta'\over b\,r^2}\,k\,\lambda\,=\,0 \ .
\label{E_fluc_eq}
\eear
Let us now write the equation of motion for the scalar field $\lambda$ in momentum space as:
\bear
\label{lambda_fluc_eq}
&& 0 = \partial_r\,\Bigg[{r^2\sin\theta h\over \sqrt{\Delta}}\,\Big[r^2\,
\Big(1\,-\,{r^2\,h\,\theta'^{\,2}\over \Delta}\Big)\lambda'+
{b\,d\,\theta'\over \sin\theta\sqrt{\Delta}}\,a_t'\Big]\Bigg] \\
&&
\quad
+b\,d\cot\theta\,a_t'\,
-k\,{d\theta'\over b\,r^2}\,E +r^2\sin\theta\,\sqrt{\Delta}\,\Big[\big(3-2b){\sin\theta\over \sqrt{\Delta}}\,+\,
{\Delta-r^2\, h\,\theta'^2\over \sin^2\theta\Delta}\,\Big]\,\lambda\rc
&&
\quad
-\omega^2\, {\sin\theta\over b^2\sqrt{\Delta}}\Big[
{(b^2+r^2\,h\,\theta'^2)(r^2\,h\,\theta'^{\,2}-\Delta)\over h\,\Delta}\,-\,
{d^2\,\theta'^{\,2}\over r^2\sin^2\theta}\,\Big]\lambda-k^2\,
{\sin\theta\sqrt{\Delta}\over b^2}\,
\Big(1-{r^2\,h\,\theta'^{\,2}\over \Delta}\Big)\lambda \ ,\nonumber
\eear
where it should be understood that $a_t'$ is given by the first equation in (\ref{at_ax_E}). The fluctuation equations (\ref{E_fluc_eq}) and (\ref{lambda_fluc_eq}) depend explicitly on the horizon radius $r_h$ through the blackening factor $h$.  This dependence can be eliminated by performing the familiar rescaling of the  radial variable 
and of the different quantities appearing in the equations. Indeed, let us rescale the radial variable $r$ and the density $d$ as in (\ref{hat_r_d}).  Moreover, we also define the rescaled frequency and momentum as:
\beq
\hat \omega\,=\,{\omega\over r_h}\,\,,
\qquad\qquad
\hat k\,=\,{k\over r_h}\,\,.
\label{hat_omega_k}
\eeq
Then, one can easily verify that the resulting equations of motion are independent of $r_h$ if the fields $E$ and $\lambda$  are rescaled appropriately.  Actually, since only the relative power of $r_h$ in these two fields matters, we can decide not to rescale the electric field $E$.  The rescaling of the scalar $\lambda$ that allows to eliminate $r_h$ is:
\beq
\hat\lambda\,=\,r_h^{2}\,\lambda\,\,.
\label{hat_lambda}
\eeq
The resulting equations of motion are just (\ref{E_fluc_eq}) and (\ref{lambda_fluc_eq}) with $r_h=1$ and with all quantities replaced by their hatted counterparts. 

The collective excitations of the brane system are dual to the quasinormal modes of the probe. The latter can be obtained by solving (\ref{E_fluc_eq}) and (\ref{lambda_fluc_eq}) for low $\omega$ and $k$ by imposing infalling boundary conditions at the horizon and the vanishing of the source terms at the UV.  At low temperature, in the so-called collisionless  quantum regime, the dominant excitation is the holographic zero sound \cite{Karch:2008fa,Karch:2009zz,Bergman:2011rf} (see also \cite{Jokela:2012se,Jokela:2012vn,Brattan:2012nb,Davison:2011ek,Kulaxizi:2008jx,Kulaxizi:2008kv}), whose dispersion relation has the form:
\beq
\hat \omega\,=\,\pm c_s\,\hat k\,-\,i\Gamma(\hat k, \hat d)\,\,.
\label{zero_sound_disp_rel}
\eeq
In (\ref{zero_sound_disp_rel}) $c_s$ is the speed of zero sound and $\Gamma$ is the attenuation. We have integrated numerically the fluctuation equations when  $\hat d$ is  large (\ie\ at low temperature) 
and we have found the value of $c_s$, both for the unflavored and the flavored backgrounds. The main conclusion from this calculation is that $c_s$ is equal to the speed of the  first sound $u_s$ (given by (\ref{speed_of_first_sound})).  As shown in Fig.~\ref{speed_of_sound}, $c_s$ reaches its maximal value ($c_2=1/\sqrt{2}$) when $m/\mu=0$, where the system is conformally invariant. In the unflavored case $c_s$ is always positive and vanishes at the quantum critical point at $\mu=m$ (see Fig.~\ref{speed_of_sound}, left). When dynamical quarks are included $c_s$ becomes imaginary when $\mu$ reaches its minimal value, which occurs when the Minkowski embeddings are thermodynamically favored.  

At higher temperature (\ie\ with small $\hat d$) the system enters into the hydrodynamic diffusive regime. The dominant mode in this case has purely imaginary frequency and a spectrum of the form:
\beq
\hat\omega\,=\,-i\hat D\,\hat k^2\,\,,
\label{hat_diffusion}
\eeq
where $\hat D$ is the rescaled diffusion constant,
\beq
\hat D\,=\,r_h\,D\,\,.
\eeq
As in the zero sound case, this dynamic calculation of the diffusion constant yields the same result as the static one. Indeed, the results obtained by numerical integration of  (\ref{E_fluc_eq}) and (\ref{lambda_fluc_eq}) coincide with the ones obtained from the Einstein relation (\ref{Diffusion_Einstein}), as shown in Fig.~\ref{Diffusion_constant_plots}.

Let us finish this section with the following observation. A careful reader would had expected some discussion on the possible instability as the WZ action has a term $C_1\wedge F\wedge F\wedge F$ which is the source of 
striping via a generic mechanism introduced in \cite{Nakamura:2009tf}. Indeed, the occurrence of tachyonic fluctuations have been confirmed in similar brane models \cite{Bergman:2011rf,Jokela:2012se}, with the subsequent construction of the striped ground state \cite{Jokela:2014dba}. In the current work, we analyzed the fluctuations of the transverse gauge field, where such an instability is expected. In this sector, one needs to analyze the coupled fluctuations of the internal gauge field $a$ and the transverse Minkowski gauge field $a_y$ at non-vanishing momentum. The corresponding equations of motion are presented in appendix \ref{app:transverse}. While we did see the precursor of the instability, a purely imaginary mode first ascending towards the upper half of the complex $\omega$ plane and then descending as a function of $k$, we were unsuccessful to finding parameter values for which case the mode would had actually become unstable. We expect that in the case in which an internal flux is turned on at the unperturbed level,  where the  contribution of the pullback of $\hat C_1$  at the background level  is non-vanishing,  the relevant WZ term can become sizable and thus implies striping in some range of parameters.

\section{Summary and outlook}
\label{summary}

In this paper we studied the phase diagram of a D6-brane probe with non-vanishing charge density in a background dual to the ABJM Chern-Simons matter theory with dynamical massless flavors at zero temperature. We analyzed the phase transition between black hole and Minkowski embeddings at zero temperature and non-vanishing chemical potential. This transition is a holographic model of a conductor-insulator phase transition between a gapless (black hole) phase and a gapped (Minkowski) phase. 

In the unflavored background we found that this transition occurs when the charge density vanishes and is of second order. Moreover, we were able to characterize the scaling behavior of the probe near the critical point. Interestingly, we found logarithmic multiplicative corrections. In the background with dynamical quarks the transition of the probe is of first order and takes place when the density is non-zero. Therefore, we have shown that, even if the change of the metric due to the backreaction to the flavor is seemingly mild, the physical effects are very important. 

It is interesting to compare our results with the one corresponding to the (2+1)-dimensional  D3-D5 intersection \cite{Karch:2007br,Ammon:2012je}. When the mass $m$ of the quarks is zero, the gravitational descriptions of both systems are equivalent and have the same thermodynamic quantities. However, for non-conformal embeddings with $m\not=0$, the ABJM probe action gets a non-trivial contribution from the Wess-Zumino term. This term is responsible for the different critical behaviors of the systems even in the absence of backreaction. 

Let us now discuss some possible extensions of our work. First of all, it would be interesting to extend our study of the Minkowski-black hole embedding phase transition  to non-zero temperature, in order to completely determine the phase diagram of the model. In the absence of the chemical potential $\mu=0$, this analysis was performed in \cite{Jokela:2012dw}. Another possible generalization would be to consider the case of massive dynamical quarks. The supergravity solution of ABJM with massive unquenched quarks at zero temperature was constructed in \cite{Bea:2013jxa}. This solution contains a scale (the mass of the sea quarks) and it would be very interesting to explore how it affects the results found here.

Turning on a suitable NSNS flat $B$ field in the ABJM supergravity solution we get the so-called ABJ background, which is dual to a Chern-Simons matter theory with gauge group $U(N+M)\times U(N)$ \cite{Aharony:2008gk}. The $B$ field breaks parity in 2+1 dimensions. The embedding of flavor brane probes in the ABJ background has been analyzed in \cite{Bea:2014yda} and the relation to the quantum Hall effect was doped out. It would be interesting to analyze possible quantum phase transitions in this ABJ system.

One of the main motivations of our work was the analysis of the effects of the dynamical quarks in the phase diagram of holographic compressible matter. We achieved this objective only partially since our backreacted background did not include the effect of the charge density on the flavor brane.  It is tempting to speculate that the smeared background at non-zero charge density would undergo a quantum phase transition similar to the one we found here.  On general grounds, one would expect having a Lifshitz geometry in the IR of such a background. Indeed, this is precisely what happens in the geometry recently found in \cite{Faedo:2015urf}, corresponding to an intersection of color D2-branes and flavor D6-branes. The study of the quantum phase transitions, as well as the collective excitations of the flavor brane, in this background is of great interest.

\vspace{1.5cm}
{\bf \large Acknowledgments}

\noindent We are grateful to  Dimitrios Zoakos for collaboration in the early stages of this work. 
We thank  Daniel Are\'an,   Georgios Itsios, and Javier Tarr\'\i o for discussions and critical readings of the manuscript.  N.~J. is supported by the Academy of Finland grant no. 1268023. 
Y.~B. and A.~V.~R. are funded by the Spanish grant FPA2014-52218-P by Xunta de Galicia (GRC2013-024), and by FEDER. Y.~B. is also supported by the Spanish FPU fellowship FPU12/00481.

\appendix

\vskip 1cm
\renewcommand{\theequation}{\rm{A}.\arabic{equation}}
\setcounter{equation}{0}

\section{More on  the background }
\label{background_appendix}

In this appendix we write in detail, following \cite{Conde:2011sw,Jokela:2012dw}, the solution of type IIA supergravity with sources that corresponds to the ABJM theory with smeared flavor branes.  Let us begin by introducing three $SU(2)$ left-invariant one-forms $\omega^i$ ($i=1,2,3$) which satisfy $d\omega^i\,=\,{1\over 2}\epsilon^{ijk}\,
\omega^i\,\omega^k$.  We will use the  $\omega^i$'s, together with a new angular coordinate $\alpha$, to parameterize the line element of the four-sphere ${\mathbb S}^4$ in (\ref{internal-metric-flavored}). We have:
\beq
ds^2_{{\mathbb S}^4}\,=\,d\alpha^2\,+\,
{\sin^2\alpha\over 4}
\left[(\omega^1)^2+(\omega^2)^2+(\omega^3)^2
\right]\ ,
\label{S4metric}
\eeq
where $0\le \alpha<\pi$. The $SU(2)$ instanton one-forms $A^i$  which fiber the ${\mathbb S}^2$ over the $ {\mathbb S}^4$ in (\ref{internal-metric-flavored}) can be written in these coordinates as:
\beq
A^{i}\,=\,-\sin^2\left({\alpha\over 2}\right)
\,\,\omega^i\,\,. 
\label{A-instanton}
\eeq
Let us next parametrize the $z^i$ coordinates of the ${\mathbb S}^2$ in (\ref{internal-metric-flavored}) by means of two angles $\theta$ and $\varphi$ ($0\le\theta<\pi$, $0\le\varphi<2\pi$), namely:
\beq
z^1\,=\,\sin\theta\,\cos\varphi\,\,,\qquad\qquad
z^2\,=\,\sin\theta\,\sin\varphi\,\,,\qquad\qquad
z^3\,=\,\cos\theta\,\,.
\label{cartesian-S2}
\eeq
Then, one can easily prove that the  $ {\mathbb S}^2$ part of the metric 
(\ref{internal-metric-flavored}) can be written as:
\beq
\left(d x^i\,+\, \epsilon^{ijk}\,A^j\,z^k\,\right)^2\,=\,(E^1)^2\,+\,(E^2)^2\ ,
\eeq
where  $E^1$ and $E^2$ are the following one-forms:
\bear
E^1&=&d\theta+\sin^2\big({\alpha\over 2}\big)\,
\left[\sin\varphi\,\omega^1-\cos\varphi\,\omega^2\right]
\rc\rc
E^2&=&\sin\theta\left[d\varphi-\sin^2\big({\alpha\over 2}\big)
\,\omega^3\right]+\sin^2\big({\alpha\over 2}\big)\,
\cos\theta\left[\cos\varphi\,\omega^1+\sin\varphi\,\omega^2\right]\ .
\label{Es}
\eear
Thus, the internal metric (\ref{internal-metric-flavored}) can be written as:
\beq
ds^2_6\,=\,{L^2\over b^2}\Big[
q^2\,d\alpha^2\,+\,{q^2\,\sin^2\alpha\over 4}
\left[(\omega^1)^2+(\omega^2)^2+(\omega^3)^2
\right]\,+\,(E^1)^2\,+\,(E^2)^2\Big]\,\,.
\eeq
The flavored ABJM background also has non-vanishing values of the RR two-forms $F_2$ and $F_4$. In order to write down their expressions, let us  first rotate the $\omega^i$'s by the two  ${\mathbb S}^2$ angles $(\theta, \varphi)$. We define three new one-forms $S^i$ ($i=1,2,3$) as:
\bear
S^1&=&\sin\varphi\,\omega^1-\cos\varphi\,\omega^2 \rc\rc
S^2&=&\sin\theta\,\omega^3-\cos\theta\left(\cos\varphi\,\omega^1+
\sin\varphi\,\omega^2\right) \rc\rc
S^3&=&-\cos\theta\,\omega^3-\sin\theta\left(\cos\varphi\,\omega^1+
\sin\varphi\,\omega^2\right)\ .
\label{rotomega}
\eear 
Next, we define the one-forms ${\cal S}^{\alpha}$   and ${\cal S}^{i}$ as:
\beq
{\cal S}^{\alpha}\,=\,d\alpha\,\,,\qquad\qquad
{\cal S}^{i}\,=\,{\sin\alpha\over 2}\,S^i \,\,,\qquad(i=1,2,3)\ ,
\label{calS}
\eeq
in terms of which the metric of the four-sphere is just 
$ds^2_{{\mathbb S}^4}=({\cal S}^{\alpha})^2+\sum_i({\cal S}^{i})^2$.   With these definitions,  we can write   the RR two-form $F_2$ for the flavored background  as:
\beq
F_2\,=\,{k\over 2}\,\,\Big[\,\,
E^1\wedge E^2\,-\,(1+\hat\epsilon)\,
\big({\cal S}^{\alpha}\wedge {\cal S}^{3}\,+\,{\cal S}^1\wedge {\cal S}^{2}\big)
\,\,\Big]\ ,
\label{F2-flavored}
\eeq
where $k$ is the Chern-Simons level.  It is important to notice that the two-form $F_2$ in (\ref{F2-flavored})  is not closed. Indeed, one can check that $dF_2\,=\,2\pi\,\,\Omega$, where $\Omega$ is the following three-form
\beq
\Omega\,=\,\hat\epsilon\,{k\over 4\pi}\,\,
\Big[\,
E^1\wedge ({\cal S}^{\xi}\wedge {\cal S}^{2}\,-\,{\cal S}^1\wedge {\cal S}^{3}\big)\,+\,
E^2\wedge ({\cal S}^{\xi}\wedge {\cal S}^{1}\,+\,{\cal S}^2\wedge {\cal S}^{3}\big)\,
\Big]\,\,,
\label{Omega}
\eeq
which does not vanish unless $\hat\epsilon=0$, \ie\ when $N_f=0$. This violation of the Bianchi identity for $F_2$ is due to the presence of delocalized flavor D6-branes ($\Omega$ is the so-called smearing form). 
The solution is completed by a constant dilaton $\phi$ given by
\beq
e^{-\phi}\,=\,{b\over 4}\,{1+\hat\epsilon+q\over 2-q}\,{k\over L}\ , 
\label{dilaton-flavored-squashings}
\eeq
and a RR four-fom $F_4$ whose expression is:
\beq
F_4\,=\,{3k\over 4}\,\,\,{(1+\hat\epsilon+q)b\over 2-q}\,\,L^2\,\,\Omega_{BH_4}\ ,
\eeq
where $\Omega_4$ is the volume form of the four-dimensional black hole (\ref{BH4-metric}). 

The flavor D6-branes are extended along the four directions of $AdS_4$ and wrap a compact three-cycle inside the internal manifold. In order to parameterize this internal cycle, let us represent the forms $\omega^i$ in terms of three angular coordinates $(\hat\theta, \hat\varphi, \hat\psi)$ as:
\bear
\omega^1 & = & \cos\hat\psi\,d\hat\theta+\sin\hat\psi\,\sin\hat\theta\,d\hat\varphi \rc
\omega^2 & = & \sin\hat\psi\,d\hat\theta-\cos\hat\psi\,\sin\hat\theta\,d\hat\varphi \rc
\omega^3 & = & d\hat\psi+\cos\hat\theta \,d\hat\varphi\,\,,
\label{w123}
\eear
with $0\le \hat\theta\le \pi$, $0\le\hat\varphi<2\pi$, $0\le\hat\psi \le 4\pi$.  The three-cycle we are looking for is topologically  ${\mathbb R}{\mathbb P}^3={\mathbb S}^3/{\mathbb Z}_2$.  It was shown in \cite{Conde:2011sw} that it can be characterized  by the conditions:
\beq
\hat\theta\, ,\ \hat\varphi \,=\,{\rm constant}\ ,
\label{RP3-cycle}
\eeq
with the coordinate $\theta$ defined in (\ref{cartesian-S2}) being a function of the radial coordinate $r$.  The induced metric on the worldvolume of the D6-brane can be written as in (\ref{induced_metric}), where $\alpha$ is the same  angle  as in  (\ref{S4metric}). The relation of the two other angles $\beta$ and $\psi$ with those introduced in (\ref{cartesian-S2}) and (\ref{w123}) is the following:
\beq
\beta\,=\,{\hat\psi\over 2}\,\,,\qquad\qquad
\psi\,=\,\varphi\,-\,{\hat\psi\over 2}\ .
\label{RP3-angles}
\eeq

\vskip 1cm
\renewcommand{\theequation}{\rm{B}.\arabic{equation}}
\setcounter{equation}{0}

\section{Fluctuation analysis}
\label{fluctuation_apendix}

Let us consider fluctuations of the gauge field $A$ and the embedding function $\theta$ as in (\ref{longitudinal_fluct_ansatz}).  We expand the induced metric $g$ and the gauge field strength as:
\beq
g\,=\,g^{(0)}\,+\,g^{(1)}\,+\,g^{(2)}\,\,,
\qquad\qquad
F\,=\,F^{(0)}\,+\,f\,\,,
\eeq
where $g^{(0)}$ is the metric written in (\ref{induced_metric}) and $F^{(0)}$ is the field strength of the unperturbed gauge connection (\ref{unperturbed_theta_A}), while $f=L^2 da$ and 
the first and second order  induced metrics 
$g^{(1)}$ and $g^{(2)}$ are given by:
\bear
&&g^{(1)}_{ij}\,d\zeta^i\,d\zeta^j\,=\,{L^2\over b^2}\,\Big[
2\,\theta' \,(\lambda'\,dr+\partial_t\lambda\,dt+\partial_x\lambda\,dx) dr\,+\,
\lambda\,\sin(2\theta)\,(d\psi+\cos\alpha \,d\beta)^2\Big] \rc
&&g^{(2)}_{ij}\,d\zeta^i\,d\zeta^j\,=\,{L^2\over b^2}\,\Big[
(\lambda'\,dr+\partial_t\lambda\,dt+\partial_x\lambda\,dx)^2+\lambda^2\,\cos (2\theta)\,
(d\psi+\cos\alpha\, d\beta)^2\Big]\,\,.
\eear
Let us next write the inverse of the zeroth-order DBI matrix $g^{(0)}+F^{(0)}$  as:
\beq
\Big(g^{(0)}+F^{(0)}\Big)^{-1}\,=\,{\cal G}^{-1}+{\cal J}\,\,,
\label{open_string_metric_def}
\eeq	
where ${\cal G}^{-1}$ is the symmetric part (the inverse open string metric) and ${\cal J}$ is the antisymmetric part. In order to write the different elements of 	${\cal G}$  and ${\cal J}$ it is quite convenient to introduce 
 the quantity $\Delta$ defined in (\ref{Delta_definition}). 
 In terms of $\Delta$, the equation for the embedding takes the form:
\beq
\partial_r\Big[{r^4\,h\,\sin\theta\over \sqrt{\Delta}}\,\theta'\Big]\,-\,r^2\,\sin\theta\cos\theta\,
\left[3-2b+{\sqrt{\Delta}\over \sin\theta}\right]\,=\,0\,\,.
\eeq
Then, the non-vanishing components of the open string metric are:
\bear
&&{\cal G}^{tt}\,=\,-{b^2+r^2\,h\,\theta'^{\,2}\over L^2\,r^2\,h\,\Delta}\,\,,
\qquad\qquad
{\cal G}^{xx}\,=\,{\cal G}^{yy}\,=\,{1\over L^2\,r^2}\,\,,\rc\rc
&&{\cal G}^{rr}\,=\,{b^2\,r^2\,h\over L^2\Delta}\,\,,
\qquad\qquad
{\cal G}^{\alpha\alpha}\,=\,{b^2\over L^2\,q}\,\,,
\qquad\qquad
{\cal G}^{\beta\beta}\,=\,{b^2\over L^2\,q\,\sin^2\alpha}\,\,,\rc\rc
&&{\cal G}^{\beta\psi}\,=\,-{b^2\cos\alpha\over L^2\,q\,\sin^2\alpha}\,\,,
\qquad\qquad
{\cal G}^{\psi\psi}\,=\,{b^2\over L^2 q}\,\Big(\cot^2\alpha\,+\,{q\over \sin^2\theta}\Big)\,\,.
\label{open_string_metric_components}
\eear
The only non-vanishing components of the antisymmetric tensor are:
\beq
{\cal J}^{tr}\,=\,-{\cal J}^{rt}\,=\,-{d\,b\over L^2\,r^2\,\sin\theta\sqrt{\Delta}}\,\,.
\eeq

At second order in the fluctuations, the DBI action is:
\bear
S_{DBI}^{(2)} & = &-T_{D6}\int d^7 \zeta e^{-\phi}\sqrt{-\det (g^{(0)}+F^{(0)})}\Bigg[{1\over 2}\Tr\big({\cal G}^{-1}g^{(2)}\big)
+{1\over  8}\Big(\Tr\big({\cal G}^{-1}g^{(1)}\big)+\Tr\big({\cal J}f\big)\Big)^2\rc
&&-{1\over 4}
\Tr\Big[({\cal G}^{-1}\,g^{(1)}\big)^2+({\cal J}\,g^{(1)}\big)^2+
4\,{\cal G}^{-1}\,g^{(1)}\,{\cal J}\,f\,+\,({\cal G}^{-1}\,f\big)^2\,+\,({\cal J}\,f\big)^2\Big]\Bigg]\,\,.
\eear
To evaluate this expression we use:
\bear
&&\Tr\big({\cal G}^{-1}\,g^{(2)}\big)\,=\,
{L^2\over b^2}\,{\cal G}^{mn}\partial_m\lambda\,\partial_n\lambda\,+\,
\big(\cot^2\theta-1)\,\lambda^2 \rc
&&\Tr\big({\cal G}^{-1}\,g^{(1)}\big)\,=\,{2L^2\over b^2}\,\theta'\,{\cal G}^{rr}\,\lambda'\,+\,2\cot\theta\,\lambda \rc
&&\Tr\Big[\big({\cal G}^{-1}\,g^{(1)}\big)^2\Big]\,=\,{4 L^4\over b^4}\,\theta'^{\,2}\,({\cal G}^{rr})^2\,
(\lambda')^2\,+\,4\,\cot^2\theta\,\lambda^2+{2L^4\over b^4}\,\theta'^{\,2}\,{\cal G}^{rr}\,
{\cal G}^{mn}\,\partial_m\lambda\partial_n\lambda\,\,,\qquad\qquad
\eear
where the indices $n$ and $m$ run over the Minkowski and radial directions. After integrating over the internal angles, we get the following second-order DBI Lagrangian:
\bear
{\cal L}^{(2)}_{DBI} & = & -{\cal N}r^2\sin\theta\sqrt{\Delta}\Bigg[{1\over 4}{\cal G}^{nm}{\cal G}^{pq}f_{mq}f_{np}+{L^2\over 2b^2}\Big(1-{L^2\over b^2}\theta'^2{\cal G}^{rr}\Big){\cal G}^{mn}\partial_m\lambda\partial_n\lambda-{1\over 2}\lambda^2 \\
&&+{L^2\over b^2}\theta'\cot\theta{\cal G}^{rr}\lambda\partial_r\lambda-{1\over 2}\Big({A_t'\theta'\over \Delta}\Big)^2(\partial_t\lambda)^2+{A_t'\theta'\over \Delta}{\cal G}^{mn}\partial_m\lambda f_{nt}+{b^2\over L^2}{A_t'\over \Delta}\cot\theta\lambda f_{tr}\Bigg] \nonumber\ .
\eear
The WZ term at second order yields the following Lagrangian density:
\beq
{\cal L}^{(2)}_{WZ}\,=\,{\cal N}\,r^2\,b\,\Big[
{r\over b}\,\cos(2\theta)\,\lambda\partial_r\lambda\,+\,
\Big(\cos(2\theta)-{r\over b}\sin(2\theta)\,\theta'\Big)\lambda^2\Big]\,\,.
\eeq
Let us now simplify these expressions. First of all, we should eliminate $A_t'$. With this purpose we notice that:
\beq
{A_t'\over \Delta}\,=\,{d\over b\,r^2\,\sin\theta\sqrt{\Delta}}\,\,.
\eeq
Secondly, we rewrite the terms with $\lambda\partial_r \lambda$ by integrating by parts and neglecting  the total derivative generated in this process. In the WZ Lagrangian we use
\beq
{r^3\over b}\,\cos(2\theta)\,\lambda\partial_r\lambda\,=\,
\Big({r^3\over b}\sin(2\theta)\,\theta'\,-\,
{3\over 2}\,{r^2\over b}\cos(2\theta)\,\Big)\,\lambda^2\,+\,
\partial_r\Big[{1\over 2}\,{r^3\over b}\,\cos(2\theta)\,\lambda^2\Big]\,\,.
\eeq
The resulting WZ Lagrangian takes the form:
\beq
{\cal L}^{(2)}_{WZ}\,=\,{\cal N}\,r^2\,b\,\big(1-{3\over 2b}\big)\,\cos(2\theta)\,\lambda^2\,\,.
\eeq
In the DBI part, we first write:
\beq
{r^4\sin\theta h\over \sqrt{\Delta}}\,\theta'\,\cot\theta\,\lambda\partial_r\lambda\,=\,
-{1\over 2}\,\partial_r\Big({r^4\sin\theta h\over \sqrt{\Delta}}\,\theta'\,\cot\theta\Big)\,\lambda^2\,+\,
\partial_r\Big({r^4\sin\theta h\over 2 \sqrt{\Delta}}\,\theta'\,\cot\theta\,\lambda^2\Big)\,\,.
\eeq
It follows that we can make the following substitution in ${\cal L}_{DBI}$:
\bear
{r^4\sin\theta h\over \sqrt{\Delta}}\,\theta'\,\cot\theta\,\lambda\partial_r\lambda
\to
-\partial_r\Big({r^4\sin\theta h\over 2 \sqrt{\Delta}}\,\theta'\cot\theta\Big)\lambda^2=
-\partial_r\Big({r^4\sin\theta h\over 2 \sqrt{\Delta}}\,\theta'\Big)\cot\theta\lambda^2+
{r^4 h\theta'^{\,2}\over 2\sin\theta\sqrt{\Delta}}\lambda^2\,\,,\rc
\eear
which, after using eq. (\ref{eom_theta_At}) for $\theta(r)$, can be written as:
\beq
{r^4\sin\theta h\over \sqrt{\Delta}}\,\theta'\,\cot\theta\,\lambda\partial_r\lambda
\to
\Bigg(\big(b-{3\over 2}\big)\,r^2\,\cos^2\theta\,+\,r^2\,
{r^2 h \theta'^{\,2}-\cos^2\theta\Delta\over 2\sin\theta\sqrt{\Delta}}\Bigg)\lambda^2\,\,.
\eeq
Taking these results into account, it is straightforward to verify that the total Lagrangian density 
${\cal L}^{(2)}={\cal L}^{(2)}_{DBI}+{\cal L}^{(2)}_{WZ}$  can be written as in (\ref{total_lag_fuct}).

\subsection{Transverse fluctuations}\label{app:transverse}
We now consider fluctuations of the gauge field along the transverse direction $y$. It turns out that these fluctuations 
are coupled to those along the internal directions. Actually, we can write the following consistent ansatz:
\beq
A\,=\,L^2\,\Big[A_t(r)\,dt\,+\,e^{-i\omega t+i kx}\,
a_y(r)\,dy\,+\,e^{-i\omega t+i kx}\,a(r)\,\big(\cos\alpha\,d\beta\,+\,d\psi\big)\Big]\,\,,
\eeq
where $a_y$ and $a$ are first-order fluctuations. The equation of motion for $a_y$ is given by:
\bear
&&\partial_r\,\Big({b^2\,r^2\,h\,\sin\theta\over \sqrt{\Delta}}\,a_y'\Big)\,+\,{\sin\theta\over r^2\,h\sqrt{\Delta}}\,
\Big[\omega^2(b^2+r^2\,h\theta'^2)-k^2\,h\,\Delta\Big]\,a_y \rc
&&\qquad\qquad
+{2i\,k\,d\,\cot\theta\over r^2}\,{b^2(2-q)\eta\over q(q+\eta)}\,\sqrt{\Delta}\,
\,a\,=\,0\,\,,
\eear
whereas that for $a$ is:
\bear
&&\partial_r\,\Big({b\,q\,r^4\,h\,\over \sin\theta\sqrt{\Delta}}\,a'\Big)\,+\,3b\,r^2\,a\,-\,{b\over q}\,r^2\sin\theta\,\sqrt{\Delta}\,a\,+
\omega^2\,{q\over b}\,{b^2+r^2\,h\,\theta'^2\over \sin\theta\,h\,\sqrt{\Delta}}\,a \rc
&&\qquad\qquad
-k^2\,{q\over b}\,{\sqrt{\Delta}\over \sin\theta}\,a\,-\,{2ikd\cot\theta \over r^2}\,{(2-q)\eta\over (q+\eta)\,b}\,\,\sqrt{\Delta}\,
a_y\,=\,0\,\,.
\eear
For our purposes, it is enough to consider the fluctuations at zero momentum ($k=0$). In this case the equation for $a_y$ is decoupled from the internal fluctuation $a$ and becomes:
\beq
\partial_r\,\Big({b^2\,r^2\,h\,\sin\theta\over \sqrt{\Delta}}\,a_y'\Big)\,+\,{\sin\theta\over r^2\,h\sqrt{\Delta}}\,
\omega^2(b^2+r^2\,h\,\theta'^2)\,a_y\,=\,0\,\,.
\label{a_y_eq_zerok}
\eeq
Explicitly, this equation for $a_y$ can be written as:
\beq
a_y''\,+\,\partial_r\,\log\Bigg[
{\sqrt{d^2+r^4\sin^2\theta}\over \sqrt{b^2+r^2\,h\,\theta'^{\,2}}}\,h\Bigg] a_y'\,+\,\omega^2\,
{b^2+r^2\,h\,\theta'^{\,2}\over b^2\,r^4\,h^2}\,a_y\,=\,0\,\,.
\label{ay_eq_k0}
\eeq
Let us expand this equation near $r=r_h$. First, we expand the embedding as in (\ref{expansion_theta_nh}):
\beq
\theta(r)\,=\,\theta_h+\theta_h'\,(r-r_h)\,+\,\cdots\,\,,
\eeq
where $\theta_h'$ is given by (see (\ref{theta_nh})):
\beq
\theta_h'\,=\,{b\,r_h\over 3}\,{\sin\theta_h\,\cos\theta_h\over d^2+r_h^4\,\sin^2\theta_h}\,
\Big[b\,r_h^2\,+\,(3-2b)\,\sqrt{d^2+r_h^4\,\sin^2\theta_h}\Big]\,\,.
\eeq
The coefficients of (\ref{ay_eq_k0})  will be expanded as:
\bear
\partial_r\,\log\Bigg[{\sqrt{d^2+r^4\sin^2\theta}\over \sqrt{b^2+r^2\,h\,\theta'^{\,2}}}\,h\Bigg] & = & {1\over r-r_h}\,+\,d_1\,\cdots \rc
\omega^2{b^2+r^2\,h\,\theta'^{\,2}\over b^2\,r^4\,h^2} & = & {A\over (r-r_h)^2}+{c_2\over r-r_h}\,+\,\cdots\,\,,
\eear
where $A$, $d_1$, and $c_2$ are:
\bear
A & = & {\omega^2\over 9\,r_h^2} \rc
d_1 & = & -{2\over r_h}\,{d^2\over d^2+r_h^4\sin^2\theta_h}\,+\,{r_h^4\,\sin\theta_h\,\cos\theta_h\over d^2+r_h^4\sin^2\theta_h}\,\theta_h'\,-\,
{3 r_h\over 2 b^2}\,\big(\theta_h')^2 \rc
c_2 & = & {\big(\theta_h')^2\over 3\, b^2\,r_h}\,\omega^2\,\,.
\eear
We now solve the equation of motion for $a_y$ in a Frobenius series around $r=r_h$ as:
\beq
a_y(r)\,=\,(r-r_h)^{\alpha}\,\big(1\,+\,\beta (r-r_h)\,+\,\cdots\big)\,\,,
\label{Frobenius_ay}
\eeq
where, for infalling boundary conditions,  the  exponent $\alpha$ is given by:
\beq
\alpha\,=\,-\,{i\omega\over 3 r_h}\,\,.
\eeq
We will also perform a low frequency expansion by considering $k\sim {\cal O} (\epsilon)$ and 
$\omega\sim { \cal O} (\epsilon^2)$. Then one can show that  $\beta\sim{ \cal O} (\epsilon^2)$ and is given by:
\beq 
\beta\,\approx\,-\alpha\,d_1\,\,.
\eeq
Let us now take the limits in opposite order. First, we consider the low frequency limit. At leading order, we can neglect the last term in (\ref{a_y_eq_zerok}) and write the equation for $a_y$ as:
\beq
\partial_r\,\Big({b^2\,r^2\,h\,\sin\theta\over \sqrt{\Delta}}\,a_y'\Big)\,=\,0\,\,.
\eeq
This equation can be immediately integrated:
\beq
a_y'\,=\,{\cal C}\,{\sqrt{\Delta}\over b^2\,r^2\,h\,\sin\theta}\,\equiv \,{{\cal C}\over G(r)}\,\,,
\eeq
where ${\cal C}$ is a constant of integration and, in the last step, we have defined the function $G(r)$. This solution can be expanded near the horizon $r=r_h$ as:
\beq
a_y'\,=\,{\cal C}\,{r_h\over 3 b\,\sqrt{d^2+r_h^4\,\sin^2\theta_h}}\,{1\over r-r_h}\,+\,\cdots\,\,.
\label{ayprime_low_nh}
\eeq
Let us now compare this near-horizon expansion with the one written in (\ref{Frobenius_ay}) for low frequency. First, we compute $a_y'$ by direct differentiation of  the expansion (\ref{Frobenius_ay}):
\beq
a_y'\,=\,\alpha (r-r_h)^{\alpha-1}\,\Big(1+\beta (r-r_h)\,+\,\cdots\Big)\,+\,(r-r_h)^{\alpha}(\beta+\cdots) \ .
\eeq
Taking into account that $\alpha\sim{\cal O}(\epsilon^2)$ and $\beta\sim{\cal O}(\epsilon^2)$, we get, at leading order in 
$\epsilon$, that:
\beq
a_y'\,=\,{\alpha \over r-r_h}\,+\,\cdots\,\,.
\label{ayprime_nh_low}
\eeq
Thus, matching (\ref{ayprime_nh_low}) and (\ref{ayprime_low_nh}), we get that the constant ${\cal C}$  is given by:
\beq
{\cal C}\,=\,{3b\over r_h}\,\sqrt{d^2+r_h^4\,\sin^2\theta_h}\,\,\alpha\,=\,
-i\,{b\over r_h^2}\,\sqrt{d^2+r_h^4\,\sin^2\theta_h}\,\omega\,\,.
\eeq
Therefore, we can write 
\beq
a_y'\,=\,-{i\over G(r)}\,{b\over r_h^2}\,\sqrt{d^2+r_h^4\,\sin^2\theta_h}\,\,\omega\,\,.
\eeq
Let us now obtain the $\left\langle J_y\,J_y\right\rangle $ correlator from these results. The term in the Lagrangian density depending on $a_y$ is given by:
\beq
{\cal L}(a_y)\,=\,-{\cal N}\,r^2\,\sin\theta\,\sqrt{\Delta}\,{\cal G}^{yy}\,{\cal G}^{rr}\,\big(f_{yr}\big)^2\,\,.
\eeq
Taking into account that $f_{ry}=L^2\,a_y'$ and that:
\beq
r^2\,\sin\theta\,\sqrt{\Delta}\,{\cal G}^{yy}\,{\cal G}^{rr}\,=\,{G(r)\over L^4}\,\,,
\eeq
we arrive at:
\beq
{\cal L}(a_y)\,=\,-{\cal F}\,\big(a_y'\big)^2\,\,,
\eeq
where ${\cal F}$ is given by:
\beq
{\cal F}\,=\,-{\cal N}\,G(r)\,\,.
\eeq
Therefore, the on-shell boundary action of $a_y$ is:
\beq
S_{{\rm on-shell}}(a_y)\,=\,\int\,d^3\,x\,\Big( {\cal F}\,a_y\,a_y'\Big)_{r\to\infty}\,\,.
\eeq
The two-point function of the transverse currents, at zero momentum, is given by:
\beq
\left\langle J_y(k)\,J_y(-k)\right\rangle\Big|_{k=0} =\Big( {\cal F}\,a_y'\Big)_{r\to\infty}\,\equiv
{\cal N}\,\Gamma_{\omega}\,i\,\omega\,\,,
\label{correlator_formula}
\eeq
where we have defined the quantity $\Gamma_{\omega}$. 
From the explicit expressions of ${\cal F}$ and $a_y$, we get:
\beq
{\cal F}\,a_y'\,=\,{\cal N}\,\,{b\over r_h^2}\,\sqrt{d^2+r_h^4\,\sin^2\theta_h}\,\,i\omega\,\,.
\eeq
Thus $\Gamma_{\omega}$ is given by:
\beq
\Gamma_{\omega}\,=\,{b\over r_h^2}\,\sqrt{d^2+r_h^4\,\sin^2\theta_h}\,\,.
\eeq
From this result we get the DC conductivity, namely:
\beq
\sigma\,=\,{\cal N}\,\Gamma_{\omega}\,=\,{\cal N}\,{b\over r_h^2}\,\sqrt{d^2+r_h^4\,\sin^2\theta_h}\,\,,
\label{dc_conductivity_ABJM}
\eeq
which is just the result written in (\ref{dc_conductivity}).

\end{document}